\documentclass[11pt,a4paper]{article}
\usepackage[utf8]{inputenc}
\usepackage[T1]{fontenc}
\usepackage{amsmath,amsthm,bm}
\usepackage[bitstream-charter]{mathdesign}
\DeclareSymbolFont{usualmathcal}{OMS}{cmsy}{m}{n}
\DeclareSymbolFontAlphabet{\mathcal}{usualmathcal}
\usepackage{graphicx}
\usepackage[top=12mm,bottom=12mm,left=30mm,right=30mm,head=12mm,includeheadfoot]{geometry}
\usepackage[expansion=false]{microtype}
\usepackage{xcolor}
\definecolor{headblue}{HTML}{1B3A5C}
\usepackage{titlesec}
\titleformat{\section}{\color{headblue}\Large\bfseries}{\thesection}{1em}{}
\titlespacing*{\section}{0pt}{1.8\baselineskip}{\baselineskip}
\usepackage[width=.90\textwidth,font=small,labelfont=bf]{caption}
\usepackage{fancyhdr}
\usepackage[colorlinks=true,linkcolor=blue!60!black,citecolor=blue!60!black,urlcolor=blue!60!black]{hyperref}
\usepackage{orcidlink}
\usepackage{doi}
\usepackage{cite}

\newtheorem{definition}{Definition}
\newtheorem{proposition}{Proposition}
\newtheorem{corollary}{Corollary}
\newtheorem{lemma}{Lemma}
\newtheorem{remark}{Remark}

\newcommand{\s}{\bm{\sigma}}
\newcommand{\x}{\bm{\xi}}
\newcommand{\e}{\bm{\eta}}
\newcommand{\E}{\mathbb{E}}
\newcommand{\Meff}{M_{\rm eff}}
\newcommand{\erf}{\mathrm{erf}}
\newcommand{\sgn}{\mathrm{sign}}

\renewenvironment{abstract}{\section*{Abstract}\bfseries}{\par}

\begin{document}
\thispagestyle{fancy}

\begin{center}{\Large\bfseries\color{headblue}
Attention over noisy examples: generalization and memorization\\ in the exponential Hopfield network
\par}\end{center}

\begin{center}\textbf{Francesco Alemanno\,\orcidlink{0000-0003-1065-2590}}\end{center}

\begin{center}
Independent researcher, supported by SolarSud T.R.E. di Miccoli Antonia, Italy (\url{https://solarsud.it})
\\[\baselineskip]
\href{mailto:francescoalemanno710@gmail.com}{\small francescoalemanno710@gmail.com}
\end{center}

\begin{abstract}
\noindent
In Hebbian learning from examples, a Hopfield network retrieves archetypes seen only through noisy examples. We study it in the exponential Hopfield model: one softmax attention step with the unlabelled examples as keys. A signal-to-noise analysis gives the overlap with an unseen archetype in closed form: a majority-rule law in which an attention entropy replaces the dataset entropy. At high temperature the archetype is lost in the crosstalk; at low temperature the softmax freezes on the closest training example, as in a random energy model, and the layer memorizes. In between, in a window widening with the square root of the network size, the layer pools the examples of the correct class and retrieves the unseen archetype as well as their majority vote allows, at any fixed load. On structured datasets, such as MNIST and fashion-MNIST, the best accuracy lies beyond the freezing.
\end{abstract}

\vspace{\baselineskip}

\section{Introduction}

Forty years after Hopfield's model \cite{Hopfield1982} and the statistical-mechanical theory of Amit, Gutfreund and Sompolinsky (AGS) \cite{AGS1985,AGS1987}, Hebbian neural networks have returned to the centre of the machine-learning stage through two independent developments. On the one hand, dense associative memories \cite{Krotov2016,Krotov2018,Demircigil2017} and their exponential limit \cite{Ramsauer2021,Lucibello2024,Albanese2025exp} have shown that a Hebbian network with sufficiently steep interactions stores an exponential number of patterns, and that the steepness of the interaction controls a transition from feature-like to prototype-like memories \cite{Krotov2016}. Moreover, one step of the zero-temperature dynamics of the exponential model is a softmax attention update with the stored patterns as keys and values \cite{Ramsauer2021,Hoover2024}, a connection also read as a nearest-neighbour or kernel-smoothing view of attention \cite{Tsai2019,Bricken2021,Millidge2022}. On the other hand, the Hebbian prescription itself has been turned from a storage rule into a learning rule: in \cite{Agliari2022concept,Alemanno2023} the network is supplied, instead of the archetypes $\x^\mu$, with $M$ corrupted examples $\e^{\mu a}$ of each of them, and the Hebbian coupling is built on the examples, without labels (unsupervised, \cite{Agliari2022concept}) or with the examples averaged class by class (supervised, \cite{Alemanno2023}). The same protocol has been solved for dense $p$-body networks in \cite{Agliari2023sup,Agliari2023unsup}, where the load becomes $\alpha_p=K/N^{p-1}$ (see also \cite{Albanese2024first,Agliari2024dreaming,Fontanari1990}).

The resulting phase diagram is controlled by the AGS load $\alpha=K/N$ and by a new parameter, the dataset entropy $\rho=(1-r^2)/(Mr^2)$, with $r$ the quality of the examples; archetypes are retrieved although they were never presented, provided $\rho$ is small enough and $\alpha$ lies below a $\rho$-dependent critical load. Two sources of noise limit the retrieval: the finite number and quality of the examples, measured by $\rho$, and the interference of the other stored patterns on the field acting on each neuron, usually called \emph{crosstalk}, which grows with the load $\alpha$. When either is too large the network enters a spin-glass phase, in which its states are correlated with none of the archetypes: in the learning setting this phase is the thermodynamic signature of overfitting.

The present work merges the two lines. We ask what happens to the $(\alpha,\rho)$ phase diagram when the examples themselves become the stored keys of the exponential Hopfield model, that is, of an attention layer, with no use of the labels. Three questions are addressed: $(i)$ whether a layer of attention, operating on a dataset of noisy examples, generalizes to the unseen archetypes and in which range of the inverse temperature $\beta$ of the softmax; $(ii)$ what is the fate of the load $\alpha$, which in the pairwise model bounds the number of archetypes that can be learnt; $(iii)$ whether the spin-glass phase has a counterpart in the exponential model, i.e.\ what overfitting looks like when the network is a transformer layer rather than a Hopfield network.

The answers are obtained analytically by a signal-to-noise analysis, which in the present setting can be carried out in closed form, and are checked by Monte Carlo simulations. We keep the dependence on $N$ explicit throughout, which is unusual in statistical mechanics but deliberate: the range of $\beta$ in which the layer generalizes is bounded from below by the crosstalk and from above by overfitting, as shown in Sec.\ \ref{sec:snr}, and the two bounds scale differently with $N$. At fixed load, the limit taken for Hebbian networks learning or storing polynomially many patterns \cite{AGS1987,Alemanno2023,Agliari2023sup,Theriault2024}, the lower bound vanishes on the scale of the upper one, while at fixed $\ln K/\sqrt N$ the contribution of each neuron to its own field takes over (Remark \ref{rem:largeK}); the exponential storage capacity is obtained at the still larger scale of fixed $\ln K/N$ \cite{Demircigil2017,Lucibello2024}, where, with examples as keys, the range is empty. Generalization thus takes place in the crossover between two limits, which the closed-form theory describes, without fitted parameters, at the sizes where simulations and attention layers actually operate (a head of dimension $N\simeq64$ attending over $K\simeq10^4$ keys has $\ln K/\sqrt N\simeq1$). In a nutshell: in \cite{Alemanno2023} the probability of reconstructing a pixel of the archetype by a majority vote over its $M$ examples is $\frac12[1+\erf(1/\sqrt{2\rho})]$, and the one-step magnetization of the network has the same error-function form; here we find that the one-step magnetization of an unseen archetype produced by an attention layer is again $m^{(1)}=\erf(1/\sqrt{2\rho_{\rm eff}})$, but the dataset entropy $\rho$ is replaced by an \emph{attention entropy} $\rho_{\rm eff}(\beta)$ that depends on how the softmax distributes its weight over the training set. At intermediate $\beta$ the load drops out ($\rho_{\rm eff}\to\rho$ for any $\alpha$); at small $\beta$ the correct class is drowned by a crosstalk with the same structure as in the pairwise model, which however grows with $K$ rather than with $\alpha$; at large $\beta$ the softmax \emph{condenses}, i.e.\ it assigns essentially all of its weight to a single training example, and the layer memorizes. Seen from the query, the logits of the softmax are Gaussian variables, independent within each class around a shift common to the class, so that the attention layer is a random energy model (REM), the system of $M$ independent Gaussian energy levels introduced by Derrida \cite{Derrida1980,Derrida1981} as the simplest spin glass: the condensation is the freezing transition of that model, located at $\beta\sqrt{N(1-r^2)}\simeq\sqrt{2\ln M}$, and it is the attention counterpart of the spin-glass phase. The window of generalization between the two edges widens with the system size as $\sqrt N$. We then test the picture on MNIST and fashion-MNIST, and on a head of the same form whose keys are trained by gradient descent, without labels, to denoise the examples.

The present model is the $p\to\infty$ end of the family of dense Hebbian networks learning from examples of \cite{Agliari2023sup,Agliari2023unsup}: there the crosstalk is still extensive in the polynomial load $\alpha_p$; here it becomes logarithmic in $K$, which is what allows the closed-form treatment. Related settings of learning from examples have been considered in \cite{Theriault2024} for dense $p$-body networks in the teacher--student scenario, in \cite{Negri2023} for the random-features Hopfield model, and in \cite{Cowsik2026} for hierarchical prototypes in polynomial dense memories. For the storage, rather than learning, properties of exponential models we refer to \cite{Demircigil2017,Lucibello2024,Albanese2025exp}, and to \cite{Agliari2026hetero} for the multilayer hetero-associative extension, whose exponential capacity and basins are obtained at $\beta\to\infty$ by a large-deviation evaluation of the noise.

\section{Setting}\label{sec:setting}

We use the vocabulary of attention throughout. The vectors stored in a network are its \emph{keys}; the state in which the network is prepared is the \emph{query}; the softmax of the overlaps between query and keys gives the \emph{attention weights}, and the state produced by one update is their weighted sum, passed through a sign.

The protocol is that of Hebbian learning from examples \cite{Agliari2022concept,Alemanno2023}. A set of binary archetypes is never shown to the network; what it receives is a few blurred examples of each, every pixel of an example agreeing with its archetype only with some probability. The examples are stored as keys, the network is prepared in a further, fresh example, and we ask what one update returns. It \emph{generalizes} when the result is close to the archetype, which it has never seen; it \emph{memorizes} when the result is close to one of its keys, i.e.\ to a training example. Three numbers control the problem: the load, i.e.\ the number of archetypes per neuron; the dataset entropy, which measures how little the examples say about their archetype; and, for the exponential network, the inverse temperature of its softmax. The networks compared below differ only in how the stored vectors enter the update: quadratically in the pairwise Hopfield networks of \cite{Agliari2022concept,Alemanno2023}, exponentially in the attention layer. Definitions \ref{def:data}--\ref{def:obs} make these notions precise.

\begin{definition}[Dataset]\label{def:data}
The archetypes are $K$ binary vectors $\x^\mu\in\{-1,+1\}^N$, $\mu=1,\dots,K$, with i.i.d.\ Rademacher entries, $\mathcal P(\xi^\mu_i=\pm1)=1/2$. The dataset $\mathcal S=\{\e^{\mu a}\}_{\mu=1,\dots,K}^{a=1,\dots,M}$ is made of $M$ examples per archetype,
\begin{equation}
\eta^{\mu a}_i=\xi^\mu_i\chi^{\mu a}_i,\qquad \mathcal P(\chi^{\mu a}_i=\pm1)=\frac{1\pm r}{2},\qquad r\in(0,1].
\end{equation}
The control parameters are the load $\alpha:=K/N$, the dataset entropy $\rho:=(1-r^2)/(Mr^2)$ \cite{Alemanno2023} and, for the exponential model, the inverse temperature $\beta$ introduced below. A \emph{query} is a further example of quality $q$, $\s^{(0)}=\x^\mu\bm\chi^{(0)}$ with $\mathcal P(\chi^{(0)}_i=\pm1)=(1\pm q)/2$, independent of $\mathcal S$; $q=1$ is the archetype itself, $q=r$ a fresh test example. The \emph{prototype} of class $\mu$ is the empirical average $\bar\eta^\mu_i:=M^{-1}\sum_a\eta^{\mu a}_i$, and $\sgn\bar\eta^\mu_i$ is its majority rule \cite{Alemanno2023}. The overlap of a state $\s$ with a vector $\bm v\in\mathbb R^N$ is $m_{\bm v}(\s):=N^{-1}\bm v\cdot\s\in[-1,1]$, and $m_\mu:=m_{\x^\mu}$ is the Mattis magnetization.
\end{definition}

\begin{definition}[Pairwise Hebbian networks \cite{Agliari2022concept,Alemanno2023}]\label{def:pair}
The Hopfield networks built on the class averages and on the examples (called supervised and unsupervised in \cite{Agliari2022concept,Alemanno2023}) have cost functions
\begin{align}
\mathcal H^{\rm sup}(\s|\mathcal S)&=-\frac{N}{2}\sum_{\mu=1}^K m_{\bar{\e}^\mu}(\s)^2,\label{eq:Hsup}\\
\mathcal H^{\rm unsup}(\s|\mathcal S)&=-\frac{N}{2M}\sum_{\mu=1}^K\sum_{a=1}^M m_{\e^{\mu a}}(\s)^2,\label{eq:Hunsup}
\end{align}
up to normalization and self-interaction terms, i.e.\ $J_{ij}\propto\sum_\mu\bar\eta^\mu_i\bar\eta^\mu_j$ and $J_{ij}\propto\sum_{\mu,a}\eta^{\mu a}_i\eta^{\mu a}_j$ respectively, with $J_{ii}=0$.
\end{definition}

\begin{definition}[Exponential Hebbian network]\label{def:exp}
Given a set of keys $\{\bm k_b\}_{b=1}^P$, the exponential Hopfield model \cite{Demircigil2017,Ramsauer2021,Lucibello2024} has cost function
\begin{equation}
\mathcal H^{\rm exp}_\beta(\s|\{\bm k_b\})=-\frac1\beta\ln\sum_{b=1}^P\exp\big[\beta N\, m_{\bm k_b}(\s)\big].
\label{eq:Hexp}
\end{equation}
Throughout, the keys are the $P=KM$ examples themselves, $\bm k_{(\mu a)}=\e^{\mu a}$, and the labels are not used; the class averages as keys, $\bm k_\mu=\bar\e^\mu$, are considered in Proposition \ref{prop:sup}.
\end{definition}

\begin{definition}[Dynamics]\label{def:dyn}
All networks evolve by the zero-temperature parallel dynamics $\sigma_i^{(t+1)}=\sgn[h_i(\s^{(t)})]$, where $h_i=-\partial\mathcal H/\partial\sigma_i$ is the post-synaptic field. For \eqref{eq:Hexp}
\begin{equation}
h_i(\s)=\sum_{b=1}^P w_b(\s)\,k_{b,i},\qquad w_b(\s)=\frac{e^{\beta N m_{\bm k_b}(\s)}}{\sum_{c}e^{\beta N m_{\bm k_c}(\s)}}.
\label{eq:attn}
\end{equation}
To first order in $\beta$, with $\langle m\rangle:=P^{-1}\sum_bm_{\bm k_b}(\s)$ and $\bar{\bm k}:=P^{-1}\sum_b\bm k_b$,
\begin{align}
w_b&\simeq\frac{1+\beta N(m_{\bm k_b}-\langle m\rangle)}{P},\nonumber\\
h_i&\simeq\bar k_i+\frac{\beta N}{P}\sum_b k_{b,i}\,(m_{\bm k_b}-\langle m\rangle).\label{eq:smallbeta}
\end{align}
\end{definition}

\begin{definition}[Observables]\label{def:obs}
Let the query $\s^{(0)}$ be an example of archetype $\x^\mu$, $\s^{(1)}$ the state after one update and $\s^{(\infty)}$ the fixed point of the iterated map, and $m^{(t)}:=m_\mu(\s^{(t)})$. The overlap with the closest training example of the same class is
\begin{equation}
n_{\max}:=\max_{a}\,m_{\e^{\mu a}}(\s^{(t)}).
\end{equation}
For the exponential network, the two statistics of the attention weights $w_b(\s^{(0)})$ of Eq.\ \eqref{eq:attn} on the query are the weight of the correct class and the participation ratio of the weights within it,
\begin{align}
W&:=\sum_{a=1}^M w_{(\mu a)}(\s^{(0)}),\label{eq:Wdef}\\
\Meff&:=\frac{\big(\sum_a w_{(\mu a)}\big)^2}{\sum_a w_{(\mu a)}^2}.\label{eq:Meffdef}
\end{align}
\end{definition}

A few comments on these definitions are in order. The prototype is the best estimate of the archetype available from the dataset, and its majority rule is what the pairwise network built on class averages computes at zero load; that network uses the labels, and it is the reference against which the attention layer is compared below.

Equation \eqref{eq:attn} is one head of self-attention with keys and values equal to the stored vectors and query $\s$, the logit being $\beta N m_{\bm k_b}=\beta\,\bm k_b\cdot\s$: one step of Definition \ref{def:dyn} is one transformer layer acting on the training set \cite{Ramsauer2021}. The expansion \eqref{eq:smallbeta} is the pairwise field built on the examples, Eq.\ \eqref{eq:Hunsup}, plus the constant $\bar k_i$ and the self-interaction. The pairwise field thus enters only at order $\beta$, on top of the mean key, which dominates as $\beta\to0$; $\beta\to\infty$ is a nearest-key lookup. The parameter $\beta$ is the inverse temperature of the Gibbs measure \eqref{eq:attn} over the keys, i.e.\ of the exponential cost function \eqref{eq:Hexp}, not of the neural dynamics, which is noiseless throughout; it is the ``$\beta$'' of \cite{Ramsauer2021} and the ``$\lambda$'' of \cite{Lucibello2024}. The logit is scaled as $\beta N m$, so that $\beta=O(1)$ is the regime of exponential capacity of \cite{Lucibello2024,Albanese2025exp}; the natural scales of the present problem will turn out to be $\beta N$ and $\beta\sqrt N$.

\begin{figure}[t]
\centering
\includegraphics[width=\textwidth]{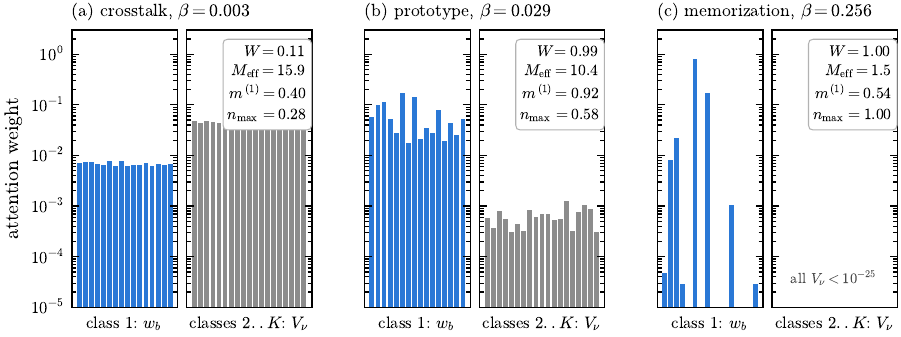}
\caption{The three regimes seen from one query. Attention weights \eqref{eq:attn} of the same query (a fresh example of archetype $\x^1$; $N=1000$, $K=20$, $M=16$, $\rho=0.2$) over the same training set at three inverse temperatures, $\beta=\beta_{\rm lo}/3$, $\sqrt{\beta_{\rm lo}\beta_{\rm hi}}$ and $3\beta_{\rm hi}$, where $\beta_{\rm lo}$ and $\beta_{\rm hi}$ are the two edges of the generalization window derived in Sec.\ \ref{sec:snr}, Eq.\ \eqref{eq:window}. In each panel the left part shows the $M$ weights $w_b$ of the keys of the correct class, the right part the total weight $V_\nu=\sum_{b\in\nu}w_b$ of each of the other $K-1$ classes (logarithmic scale, shared; bars below $10^{-5}$ are not drawn). $W=\sum_aw_{1a}$, Eq.\ \eqref{eq:Wdef}, is the sum of the blue bars and $\Meff$, Eq.\ \eqref{eq:Meffdef}, their participation ratio; $m^{(1)}$ and $n_{\max}$ are the overlaps of the updated state with the archetype and with the closest example. (a) Crosstalk: the correct class weighs as much as any other. (b) Prototype: the correct class carries all the weight, spread over its examples. (c) Memorization: one example carries it, and is returned.}
\label{fig:regimes}
\end{figure}

The one-step state $\s^{(1)}$ is what a transformer layer computes, the fixed point $\s^{(\infty)}$ is the Hopfield retrieval. Generalization is measured by $m^{(t)}$, which equals one when the state is the archetype; memorization by $n_{\max}$, which equals one when the state is a training example. $W\in[0,1]$ is the total weight that the softmax assigns to the examples of the correct class. $\Meff\in[1,M]$ is the participation ratio of the weights within that class, i.e.\ the number of examples that effectively contribute to the weighted sum: $\Meff=M$ when the $M$ examples carry equal weight (the layer \emph{pools} them all, and the update is the majority rule), $\Meff=1$ when a single example carries all the weight (the layer has \emph{condensed} on it). We refer to $W$ and $\Meff$ as the attention statistics. Condensation is thus a property of the weights ($\Meff=O(1)$), memorization one of the output ($n_{\max}\to1$): with keys equal to examples of a single archetype the two coincide, on structured data they need not (Sec.\ \ref{sec:mnist}). Figure \ref{fig:regimes} shows them on one query, at three temperatures, together with the outcome of the update: it anticipates the three regimes derived in the next section.

\section{Signal-to-noise analysis}\label{sec:snr}

The signal-to-noise technique \cite{AGS1985,Coolen2005,Alemanno2023} looks at the field acting on one neuron when the network is prepared in the query, treats it as a sum of many independent contributions, and replaces its distribution over the quenched disorder (archetypes, examples and query noise) by a Gaussian. The fraction of neurons aligned with the archetype after one update then follows from the ratio between the mean of the field, the \emph{signal}, coming from the keys of the queried archetype, and its fluctuations, dominated by the \emph{crosstalk} of all the other keys.

For the pairwise network this gives back, for a query of any quality, the law of \cite{Alemanno2023}. For the attention layer the field is a weighted vote of the examples, and the weights depend on the query. The key observation is that, seen from the query, the overlaps of the keys with it are Gaussian: those of the correct class share a common positive shift, those of each wrong class a common random one, and within every class they spread by the same amount. The logits are therefore the energy levels of a random energy model \cite{Derrida1980,Derrida1981} with a planted block, and the softmax is its Gibbs measure. Two numbers then decide the quality of the vote: how much weight goes to the correct class, and over how many of its examples it is spread. The first is a ratio of Boltzmann sums, computed by averaging numerator and denominator separately; the second is the participation ratio of a REM. A last ingredient is the term of the very neuron that is being updated, which sits inside every logit and acts on that neuron as an extra field towards the query. The statements that follow carry out these steps in this order; we return to their meaning after Proposition \ref{prop:sup}.

\begin{proposition}[pairwise network built on class averages]\label{prop:pair}
For the network \eqref{eq:Hsup} prepared in a query of quality $q$ of archetype $\x^1$, in the limit $N\to\infty$ at fixed $\alpha,\rho$ and $M\gg1$,
\begin{equation}
m^{(1)}_{\rm pair}=\erf\!\left(\frac{1}{\sqrt{2\rho+2\alpha(1+\rho)^2/q^2}}\right).
\label{eq:pair}
\end{equation}
\end{proposition}
\begin{proof}
With the normalization $J_{ij}=\frac{1}{N\mathcal R}\sum_\mu\bar\eta^\mu_i\bar\eta^\mu_j$, $\mathcal R=r^2+(1-r^2)/M$, the field splits into the signal $\bar\eta^1_i\, m_{\bar\e^1}(\s^{(0)})/\mathcal R$ and the crosstalk $\sum_{\mu>1}\bar\eta^\mu_i m_{\bar\e^\mu}(\s^{(0)})/\mathcal R$. By the law of large numbers $m_{\bar\e^1}(\s^{(0)})\to rq$, while $\bar\eta^1_i=\xi^1_i(r+\zeta_i\sqrt{(1-r^2)/M})$ with $\zeta_i$ standard Gaussian for $M\gg1$, so that the signal, multiplied by $\xi^1_i$, has mean $\mathcal M=r^2q/\mathcal R$ and variance $(rq)^2(1-r^2)/(M\mathcal R^2)$. Each crosstalk term is a product of two independent centred variables of variances $\mathcal R$ and $\mathcal R/N$, and there are $K-1$ of them, giving variance $\alpha$ (we exclude $J_{ii}$). Dividing the total variance $\mathcal V$ by $\mathcal M^2$ one gets $\mathcal V/\mathcal M^2=\rho+\alpha(1+\rho)^2/q^2$, whence \eqref{eq:pair}. (The condition $M\gg1$ is used only for the Gaussianity of $\zeta_i$; the crosstalk is Gaussian by the central limit theorem over $\mu$.) For $q=1$, \eqref{eq:pair} is Eq.\ (68) of the Supplementary Material of \cite{Alemanno2023}.
\end{proof}

\begin{lemma}[logit statistics]\label{lem:logits}
Let $\s^{(0)}$ be a query of quality $q$ of $\x^1$. Conditionally on the query and on the archetypes, the overlaps of the keys with it are, for $N\gg1$, Gaussian variables
\begin{align}
m_{\e^{1a}}(\s^{(0)})&\simeq qr+z_a\sqrt{(1-r^2)/N},\qquad a=1,\dots,M,\label{eq:logit1}\\
m_{\e^{\mu b}}(\s^{(0)})&\simeq rg_\mu+z_{\mu b}\sqrt{(1-r^2)/N},\qquad \mu>1,\label{eq:logit2}
\end{align}
where the $z$'s are independent standard Gaussians and
\begin{equation*}
g_\mu:=N^{-1}\sum_i\xi^\mu_i\xi^1_i\chi^{(0)}_i\simeq\mathcal N(0,1/N)
\end{equation*}
is a shift shared by the $M$ keys of class $\mu$. The logits $\beta Nm$ are thus a REM whose $M$ levels of the correct class have mean $\beta Nqr$, those of a wrong class $\mu$ a random mean $\beta Nrg_\mu$ of spread $\beta r\sqrt N$ across classes, and the levels of every class the same spread $x:=\beta\sqrt{N(1-r^2)}$.
\end{lemma}
\begin{proof}
Conditionally on $\chi^{(0)}$, $Nm_{\e^{1a}}=\sum_i\chi^{1a}_i\chi^{(0)}_i$ is a sum of independent $\pm1$ variables with means $r\chi^{(0)}_i$ and variances $1-r^2$; its mean is $r\sum_i\chi^{(0)}_i=Nrq+O(\sqrt N)$, the $O(\sqrt N)$ term being the same for all $a$. For $\mu>1$, $\xi^\mu_i\xi^1_i\chi^{\mu b}_i\chi^{(0)}_i$ has conditional mean $r\xi^\mu_i\xi^1_i\chi^{(0)}_i$ and variance $1-r^2$, whence \eqref{eq:logit2}. Averaged also over the query and the archetypes, the unconditional variances are $(1-q^2r^2)/N$ and $1/N$, but it is the conditional spread that governs the softmax, since the shifts cancel in the normalized weights of a class.
\end{proof}

\begin{proposition}[weight of the correct class]\label{prop:W}
In the annealed approximation, in which the ratio of two random sums is replaced by the ratio of their averages, valid for $x\lesssim\sqrt{2\ln M}$ and $\beta r\sqrt N\lesssim\sqrt{2\ln K}$,
\begin{equation}
W(\beta)=\Big[1+(K-1)\exp\!\big(-N\beta qr+\tfrac12N\beta^2r^2\big)\Big]^{-1}.
\label{eq:W}
\end{equation}
\end{proposition}
\begin{proof}
Averaging the Boltzmann sums over the Gaussian variables of Lemma \ref{lem:logits},
\begin{align*}
\E\sum_a e^{\beta Nm_{\e^{1a}}}&=Me^{\beta Nqr+x^2/2},\\
\E\sum_{\mu>1,b}e^{\beta Nm_{\e^{\mu b}}}&=(K-1)Me^{\beta^2Nr^2/2+x^2/2};
\end{align*}
the ratio of the former to the total gives \eqref{eq:W}. The annealed average is accurate while each sum is dominated by typical levels, i.e.\ below the freezing point of the corresponding REM, which gives the stated range; beyond it the sums are dominated by their largest terms and the exponent in \eqref{eq:W} is no longer the right one (it would turn non-monotonic at $\beta=q/r$).
\end{proof}

\begin{proposition}[condensation of the softmax]\label{prop:rem}
Let
\begin{equation*}
\Phi_M(x):=\frac1M\,\E\!\left[\frac{\big(\sum_{a=1}^Me^{xz_a}\big)^2}{\sum_{a=1}^Me^{2xz_a}}\right],
\end{equation*}
with $z_a$ i.i.d.\ standard Gaussian, be the participation ratio (normalized to one) of the Gibbs measure of a REM with $M$ levels at inverse temperature $x$. Then
\begin{equation}
\E\,\Meff=M\,\Phi_M(x),\qquad x=\beta\sqrt{N(1-r^2)},
\label{eq:Meff}
\end{equation}
independently of the quality $q$ of the query, the average being over the disorder of Lemma \ref{lem:logits}, and the same holds within each wrong class, $\E\,\Meff'=\E\,\Meff$. For small $x$,
\begin{equation*}
\Phi_M(x)=1-(1-M^{-1})x^2+O(x^4);
\end{equation*}
for $M\to\infty$ the measure freezes at $x_c=\sqrt{2\ln M}$, and in the frozen phase $\E\sum_aw_a^2=1-x_c/x$ \cite{Derrida1980,Derrida1981,Derrida1985b,MPV1985}, so that
\begin{equation*}
M\Phi_M(x)=\E\Big[\frac{1}{\sum_aw_a^2}\Big]\ge\Big(1-\frac{x_c}{x}\Big)^{-1}.
\end{equation*}
\end{proposition}
\begin{proof}
Within the correct class, the common factor $e^{\beta Nqr}$ cancels from the normalized weights, which become $w_a\propto e^{xz_a}$ by Lemma \ref{lem:logits}; the average participation ratio of $M$ such weights is by definition $M\Phi_M(x)$, and the levels of a wrong class have the same spread ($\Meff$ itself fluctuates from query to query: its standard deviation is $2.1$ at $x=1$ for $M=16$). The small-$x$ expansion follows from $e^{xz}=1+xz+x^2z^2/2+\dots$ (Appendix B); the freezing point is where the entropy $\ln M$ of the levels equals the energy gain $x^2/2$ of the extreme level, the frozen-phase value of $\E\sum_aw_a^2$ is the classical result of \cite{Derrida1985b,MPV1985}, and the bound is Jensen's inequality for the convex function $y\mapsto1/y$. It is strict, because the weights remain random in the frozen phase (they have the Poisson--Dirichlet statistics of \cite{MPV1985}); Appendix B quantifies the gap.
\end{proof}

The logit of every key contains the term $\beta k_{b,i}\sigma_i$ of the neuron $i$ being updated; we call it the \emph{self-coupling}, the exponential-model analogue of the diagonal term $J_{ii}$ of the pairwise model, which is removed there by convention (Definition \ref{def:pair}) but is part of the softmax here and is kept.

\begin{figure}[t]
\centering
\includegraphics[width=\textwidth]{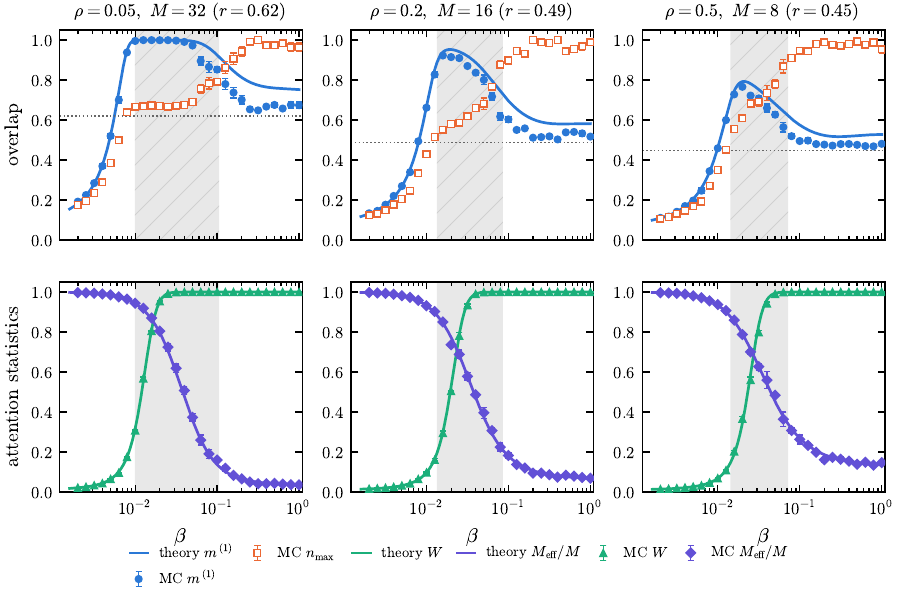}
\caption{One attention step over the training set as a function of the inverse temperature $\beta$, for $N=1000$, $\alpha=0.1$ and three datasets of increasing entropy (columns: $(\rho,M)=(0.05,32)$, $(0.2,16)$, $(0.5,8)$, i.e.\ $r=0.62$, $0.49$, $0.45$). Top: magnetization of the unseen archetype, theory \eqref{eq:main} (line) versus Monte Carlo (full circles), and overlap $n_{\max}$ with the closest training example (open squares); the dotted line is the quality $r$ of a single example. Bottom: weight of the correct class $W$ and pooled fraction $\Meff/M$, theory \eqref{eq:W}, \eqref{eq:Meff} (lines) versus Monte Carlo (symbols). The grey band is the analytic window \eqref{eq:window}. The three regimes of Corollary \ref{cor:regimes} are visible from left to right: crosstalk ($W\ll1$), prototype (crosstalk below $\rho$, $\Meff\simeq M$), memorization ($\Meff\to1$, $n_{\max}\to1$, $m^{(1)}\to r$).}
\label{fig:lambda}
\end{figure}

\begin{proposition}[one attention step]\label{prop:main}
For the exponential network \eqref{eq:Hexp} with keys $=$ examples, prepared in a query of quality $q$ of $\x^1$, in the Gaussian signal-to-noise approximation of Proposition \ref{prop:pair} (the field is replaced by a Gaussian of the same mean and variance),
\begin{equation}
m^{(1)}=\erf\!\left(\frac{1}{\sqrt{2\rho_{\rm eff}(\beta)}}\right),
\label{eq:main}
\end{equation}
with the attention entropy
\begin{align}
\rho_{\rm eff}&=\frac{\mathcal V}{\big(rW+\tau q\big)^2},\label{eq:rhoeff}\\
\mathcal V&=\frac{(1-r^2)W^2}{\Meff}+\frac{(1-W)^2}{K-1}\Big[r^2+\frac{1-r^2}{\Meff'}\Big]+\tau^2(1-q^2),\nonumber
\end{align}
with $\tau=\tanh\beta$, $W$ given by \eqref{eq:W} and $\Meff=\Meff'$ replaced by their average \eqref{eq:Meff}, as long as the field is a sum of many comparable terms, i.e.\ $\Meff\gg1$ or $W\ll1$, and the self-coupling is small compared with the rest of the field (Remark \ref{rem:limits}).
\end{proposition}
\begin{proof}
Write the field \eqref{eq:attn} as
\begin{equation*}
h_i=\sum_aw_{1a}\eta^{1a}_i+\sum_{\mu>1,b}w_{\mu b}\eta^{\mu b}_i
\end{equation*}
and multiply by $\xi^1_i$. The $i$-th factor of each Boltzmann weight,
\begin{equation*}
e^{\beta k_{b,i}\sigma_i}=\cosh\beta+\sigma_ik_{b,i}\sinh\beta,
\end{equation*}
is singled out: writing $h^{\setminus i}_i:=\sum_bw'_bk_{b,i}$ with weights $w'_b$ that do not contain site $i$, one finds
\begin{equation*}
h_i=\frac{h^{\setminus i}_i+\tau\sigma_i}{1+\tau\sigma_ih^{\setminus i}_i},\qquad \sgn h_i=\sgn\big(h^{\setminus i}_i+\tau\sigma^{(0)}_i\big),
\end{equation*}
the denominator being positive, so that the second identity is exact: the self-coupling enters the update as an additive field $\tau\sigma^{(0)}_i$, and the weights $w'_b$ are described by Lemma \ref{lem:logits}. Conditionally on the weights, $\xi^1_i\eta^{1a}_i=\chi^{1a}_i$ has mean $r$ and variance $1-r^2$, so the correct-class term has mean $rW$ and variance $(1-r^2)\sum_aw_{1a}^2$, i.e.\ $(1-r^2)W^2/\Meff$. For a wrong class $\mu$, $\xi^1_i\eta^{\mu b}_i=\xi^1_i\xi^\mu_i\chi^{\mu b}_i$ and all its examples share the factor $\xi^1_i\xi^\mu_i$: the class contributes $\xi^1_i\xi^\mu_i\sum_bw_{\mu b}\chi^{\mu b}_i$, of zero mean and variance $r^2V_\mu^2+(1-r^2)\sum_bw_{\mu b}^2$ with $V_\mu=\sum_bw_{\mu b}$. Summing over the $K-1$ wrong classes with
\begin{equation*}
V_\mu\simeq\frac{1-W}{K-1},\qquad \sum_bw_{\mu b}^2\simeq\frac{V_\mu^2}{\Meff'},
\end{equation*}
gives the second term of the numerator; the self-coupling $\tau\xi^1_i\sigma^{(0)}_i=\tau\chi^{(0)}_i$ has mean $\tau q$ and variance $\tau^2(1-q^2)$. Applying the central limit theorem to $\xi^1_ih_i$ and using $m^{(1)}=\E\,\sgn(\xi^1_ih_i)$ yields \eqref{eq:main} with $\rho_{\rm eff}=\mathcal V/\mathcal M^2$; the Gaussian step requires $N\to\infty$ for the crosstalk and $M\gg1$ for the within-class vote, as in Proposition \ref{prop:pair} (the exact $M$-term majority vote differs from \eqref{eq:main} by $0.004$ at $M=16$, $\rho=0.2$). The shifts $g_\mu$ of Lemma \ref{lem:logits} make the $V_\mu$ unequal: in the annealed approximation the $r^2$ part of the crosstalk is multiplied by $e^{\beta^2Nr^2}[1+(e^{x^2}-1)/M]$, as a Monte Carlo check confirms (Appendix A); the factor is $1.04$--$1.15$ between $\beta_{\rm lo}$ and $\beta^*$ in Fig.\ \ref{fig:lambda} and changes $m^{(1)}$ by less than $0.01$, and at larger $\beta$ the crosstalk is exponentially small. Finally, \eqref{eq:rhoeff} uses the average \eqref{eq:Meff} in place of the fluctuating $\Meff$; using $\E[1/\Meff]$ instead lowers $m^{(1)}$ appreciably only where the Gaussian approximation is already inaccurate, and in the direction of the Monte Carlo (Appendix B).
\end{proof}

\begin{remark}[condensed limit]\label{rem:cond}
For $\Meff\to1$ the field is $\eta^{1a^*}_i+\tau\sigma^{(0)}_i$ with $\tau<1$, so that $\s^{(1)}=\e^{1a^*}$, the example closest to the query, and $m^{(1)}\to r$ up to the selection bias of $a^*$; Eq.\ \eqref{eq:main} tends instead to $\erf\big(r/\sqrt{1-r^2}\big)$.
\end{remark}

\begin{corollary}[three regimes and the window]\label{cor:regimes}
Let $q=r$, $K,M\gg1$ and
\begin{equation*}
K\ll K_\times:=\frac{1+\rho}{(M+2)\rho}\Big(\frac{2Nr^2}{\ln K}\Big)^2,
\end{equation*}
so that the self-coupling can be neglected (Appendix D); $K_\times$ grows as $N^2/\ln^2K$, and the condition holds at any fixed load. Then:
\begin{itemize}\setlength\itemsep{1pt}
\item[(i)] \emph{Crosstalk regime}, $2\beta Nr^2\lesssim\ln K$: $W\ll1$, $\Meff\simeq\Meff'\simeq M$ and
\begin{equation}
\rho_{\rm eff}\simeq\rho+\frac{1+\rho}{K-1}\,\frac{(1-W)^2}{W^2};
\label{eq:crosstalk}
\end{equation}
for $\beta\to0$, $W\to1/K$ and the second term tends to $(K-1)(1+\rho)$.
\item[(ii)] \emph{Prototype regime}, $\ln K\lesssim2\beta Nr^2$ and $x\lesssim1$: the second term of \eqref{eq:crosstalk} is below $\rho$ and decays as $e^{-2N\beta r^2}$, $\Meff\simeq M$, and $\rho_{\rm eff}\simeq\rho$, $m^{(1)}\simeq\erf(1/\sqrt{2\rho})$ independently of $\alpha$.
\item[(iii)] \emph{Memorization regime}, $x\gtrsim\sqrt{2\ln M}$: $W\simeq1$, $\E\,\Meff=M\Phi_M(x)=O(1)$, and Remark \ref{rem:cond} applies for $x\gg x_c$, where $\Meff\to1$.
\end{itemize}
The window is bounded by the \emph{crosstalk edge} $\beta_{\rm lo}$, where the second term of \eqref{eq:crosstalk} equals $\rho$, and by the \emph{condensation edge} $\beta_{\rm hi}$, where $x=\sqrt{2\ln M}$. With $(1-W)/W=(K-1)e^{-N\beta r^2}$ from \eqref{eq:W}, whose quadratic term is negligible at these $\beta$,
\begin{equation}
\beta_{\rm lo}=\frac{\ln\frac{(K-1)(1+\rho)}{\rho}}{2Nr^2}<\beta<\beta_{\rm hi}\simeq\frac{\sqrt{2\ln M}}{\sqrt{N(1-r^2)}},
\label{eq:window}
\end{equation}
or, in terms of the mean $\bar m$ and the conditional standard deviation $\sigma_m$ of the within-class overlaps of Lemma \ref{lem:logits} ($\bar m=r^2$, $\sigma_m=\sqrt{(1-r^2)/N}$, $\rho=N\sigma_m^2/(M\bar m)$),
\begin{equation}
\beta_{\rm lo}=\frac{\ln[(K-1)(1+\rho)/\rho]}{2N\bar m},\qquad \beta_{\rm hi}\simeq\frac{\sqrt{2\ln M}}{N\sigma_m}.
\label{eq:window_emp}
\end{equation}
At the crosstalk edge, to leading order in $x$, $\rho_{\rm eff}\simeq2\rho$ and
\begin{equation*}
W(\beta_{\rm lo})=\Big[1+\sqrt{(K-1)\rho/(1+\rho)}\Big]^{-1}\simeq\sqrt{\frac{1+\rho}{K\rho}}.
\end{equation*}
The width of the window is
\begin{equation*}
\ln\frac{\beta_{\rm hi}}{\beta_{\rm lo}}=\frac12\ln N+\ln\frac{2r^2\sqrt{2\ln M}}{\sqrt{1-r^2}\,\ln[(K-1)(1+\rho)/\rho]},
\end{equation*}
and, within the approximations defining the two edges, the window is non-empty when
\begin{equation}
\ln\frac{(K-1)(1+\rho)}{\rho}\lesssim 2r^2\sqrt{\frac{2N\ln M}{1-r^2}}.
\label{eq:Kmax}
\end{equation}
\end{corollary}
\begin{proof}
With $\tau=0$, $q=r$ and $\Meff=\Meff'=M$, dividing numerator and denominator of \eqref{eq:rhoeff} by $r^2W^2$ and using $(1-r^2)/(Mr^2)=\rho$, $[r^2+(1-r^2)/M]/r^2=1+\rho$ gives \eqref{eq:crosstalk}. By \eqref{eq:W},
\begin{equation*}
(1-W)/W=(K-1)e^{-N\beta r^2+N\beta^2r^2/2},
\end{equation*}
so the second term of \eqref{eq:crosstalk} equals
\begin{equation*}
(1+\rho)(K-1)e^{-2N\beta r^2+N\beta^2r^2};
\end{equation*}
it decreases from $(1+\rho)(K-1)$ at $\beta=0$, where $W=1/K$, to $O(1)$ at $2N\beta r^2=\ln K$, where $W\simeq K^{-1/2}$, while $\Meff=\Meff'=M\Phi_M(x)\simeq M$ because $x=O(\ln K/\sqrt N)$ there (Proposition \ref{prop:rem}); this is (i). Above, the same term keeps decaying as $e^{-2N\beta r^2}$, $W\to1$, and
\begin{equation*}
\Meff=M[1-(1-M^{-1})x^2+O(x^4)]
\end{equation*}
by the small-$x$ expansion of Proposition \ref{prop:rem}, so that $\rho_{\rm eff}\to\rho$ up to a pooling loss of relative order $x^2$; $K$ enters \eqref{eq:crosstalk} only through that decaying term, whence (ii). For $x\gtrsim\sqrt{2\ln M}$, i.e.\ $\beta\gtrsim\beta_{\rm hi}>\beta_{\rm lo}$ when the window is non-empty, the weight of the wrong classes is exponentially small in $N\beta$ (beyond the annealed range of \eqref{eq:W} the largest logit still belongs to the correct class, whose levels are shifted by $N\beta r^2\gg\beta r\sqrt{2N\ln K}$, the extreme shift of a wrong class, since $\ln K\ll Nr^2$), so that $W\simeq1$, while the measure within the class freezes, $\E\,\Meff=M\Phi_M(x)=O(1)$, and for $x\gg x_c$ the field reduces to that of Remark \ref{rem:cond}; this is (iii).

Setting the second term of \eqref{eq:crosstalk} equal to $\rho$ gives $(1-W)/W=\sqrt{\rho(K-1)/(1+\rho)}$, i.e.\ the stated $W(\beta_{\rm lo})$; equating it to $(K-1)e^{-N\beta r^2}$ gives $\beta_{\rm lo}$ in \eqref{eq:window}, and $x=\sqrt{2\ln M}$ gives $\beta_{\rm hi}$. The quadratic term dropped from the exponent of \eqref{eq:W} is
\begin{equation*}
N\beta_{\rm lo}^2r^2/2=\ln^2[(K-1)(1+\rho)/\rho]/(8Nr^2),
\end{equation*}
a fraction $\beta_{\rm lo}/2$ of the linear term, negligible when $\ln K\ll Nr^2$, i.e.\ $\beta_{\rm lo}\ll1$, which also keeps $\beta_{\rm lo}$ within the range of validity of \eqref{eq:W}. Equation \eqref{eq:window_emp} is the substitution $r^2=\bar m$, $(1-r^2)/N=\sigma_m^2$. At the edge
\begin{equation*}
x_{\rm lo}=\beta_{\rm lo}\sqrt{N(1-r^2)}=O(\ln K/\sqrt N),
\end{equation*}
so the first term of \eqref{eq:crosstalk} is $\rho$ to leading order in $x$ and $\rho_{\rm eff}\simeq2\rho$. The width is $\ln(\beta_{\rm hi}/\beta_{\rm lo})$, and \eqref{eq:Kmax} is the condition $\beta_{\rm lo}<\beta_{\rm hi}$. Finally, the self-coupling $\tau=\tanh\beta$ competes with the signal $rW$ only where $W$ is small, i.e.\ at the crosstalk edge: Appendix D solves $\rho_{\rm eff}=2\rho$ with the full expression \eqref{eq:rhoeff} and finds that $\tau\simeq\beta_{\rm lo}$ leaves $W(\beta_{\rm lo})$ unchanged as long as $(M+2)\rho\tau^2\ll(1+\rho)/(K-1)$, which with $\beta_{\rm lo}\simeq\ln K/(2Nr^2)$ is $K\ll K_\times$; at fixed load $K=\alpha N$, $K/K_\times=O(\ln^2N/N)$.
\end{proof}

\begin{remark}[very large $K$]\label{rem:largeK}
For $K\gg K_\times$ the self-coupling cannot be neglected at the crosstalk edge: the noise $\tau^2(1-q^2)$ of the query, which the pooling does not reduce, sets the edge, the correct class must carry a weight of order $\tau\sqrt M$ rather than $K^{-1/2}$, and $\beta_{\rm lo}\simeq\ln K/(Nr^2)$, twice \eqref{eq:window} (Appendix D). For $K=\exp(c\sqrt N)$ the window is then non-empty for
\begin{equation*}
c<c_{\rm crit}=r^2\sqrt{2\ln M/(1-r^2)},
\end{equation*}
half the constant that \eqref{eq:Kmax} would give. There
\begin{equation*}
x_{\rm lo}=(c/c_{\rm crit})\sqrt{2\ln M}=O(1);
\end{equation*}
$c_{\rm crit}$ bounds the selection of the correct class, while the pooling $\Meff/M=\Phi_M(x_{\rm lo})$, and with it $m^{(1)}$, stays below the majority-rule value by a finite amount; only at fixed load does $x_{\rm lo}\to0$.
\end{remark}

\begin{proposition}[optimal temperature]\label{prop:opt}
Inside the window, neglecting the self-coupling and to leading order in $x$, $\rho_{\rm eff}-\rho$ is the sum of the pooling loss $\rho(1-M^{-1})x^2$ and of the crosstalk $(1+\rho)(K-1)e^{-2N\beta r^2}$, and at the maximum $\beta^*$ of $m^{(1)}$
\begin{equation}
\text{pooling loss}=N\beta^*r^2\times\text{crosstalk},\qquad N\beta^*r^2\simeq\ln K;
\label{eq:balance}
\end{equation}
$\beta^*$ is given in closed form by Eq.\ \eqref{eq:betastar} of Appendix C, and, within the same approximation, the weight of the correct class at the maximum decreases with the load as $(\alpha\ln K)^{-1/2}$.
\end{proposition}
\begin{proof}
The pooling loss follows from \eqref{eq:rhoeff} with $\Meff=M\Phi_M(x)$ and the small-$x$ expansion of Proposition \ref{prop:rem}, the crosstalk from \eqref{eq:crosstalk} with \eqref{eq:W}; the first grows as $\beta^2$, the second decays exponentially, and differentiating their sum gives \eqref{eq:balance} (Appendix C).
\end{proof}

\begin{proposition}[class averages as keys]\label{prop:sup}
With the $K$ class averages $\bar\e^\mu$ as keys, the factorization of the self-coupling used in the proof of Proposition \ref{prop:main} does not apply, since the keys are not binary; we neglect the bias $e^{\beta\bar\eta^\mu_i\sigma_i}$ of the $i$-th factor of each weight, which is of relative size $\beta r$ and cancels between numerator and denominator of \eqref{eq:attn} as soon as one key carries the weight. Then $m^{(1)}=\erf(1/\sqrt{2\rho_{\rm eff}})$ with
\begin{equation*}
\rho_{\rm eff}=\frac{1}{r^2W^2}\left[\frac{(1-r^2)W^2}{M}+\frac{(1-W)^2}{K-1}\Big(r^2+\frac{1-r^2}{M}\Big)\right],
\end{equation*}
$W$ being the weight of the correct key; for $\beta\to\infty$, $\rho_{\rm eff}\to\rho$.
\end{proposition}

We now read these results, starting from the main formula. Equation \eqref{eq:main} has the form of the majority-rule bound $\frac12[1+\erf(1/\sqrt{2\rho})]$ of \cite{Alemanno2023}, recalled in the Introduction: the attention layer performs a \emph{weighted} majority rule, and $\rho_{\rm eff}$ is the entropy of that weighted vote.\footnote{We keep the name ``entropy'' for $\rho$ and $\rho_{\rm eff}$ as in \cite{Alemanno2023}, where it is justified by the monotonic relation between $\rho$ and the conditional entropy of the archetype given the examples; $\rho_{\rm eff}$ is a noise-to-signal ratio of the weighted vote and should not be confused with the Shannon entropy of the attention weights, also called attention entropy in the machine-learning literature \cite{Zhai2023}.} Its three terms are the noise of the pooled examples of the correct class, the crosstalk of the wrong classes, and the self-coupling, which biases the update towards the query. For the pairwise network the vote is unweighted and the crosstalk is of order $\alpha$: the archetypes are stable under one update only in the region $\alpha(1+\rho)^2+\rho\lesssim1$, and a query of lower quality $q<1$ shrinks that region by the factor $q^2$ in front of the load.

Corollary \ref{cor:regimes} sorts the temperatures by which term of $\rho_{\rm eff}$ dominates. At small $\beta$ the vote of the correct class is drowned by the $K-1$ wrong ones as long as $W\lesssim K^{-1/2}$, through a crosstalk term with the same structure as $\alpha(1+\rho)^2/q^2$ in \eqref{eq:pair}; as $\beta\to0$ the limit is the mean of all keys, controlled by $K$ rather than by $\alpha$; it is not the pairwise network. At intermediate $\beta$ the crosstalk falls below the dataset entropy and keeps decaying exponentially in $N\beta$ ($W$ rises from $\sim K^{-1/2}$ to one) while the layer still pools all the examples of the class: $\rho_{\rm eff}\to\rho$, and the retrieval boundary in the $(\alpha,\rho)$ plane is a horizontal line. At large $\beta$ the pooling degrades gradually for $1\lesssim x\lesssim\sqrt{2\ln M}$, where \eqref{eq:main} with $\Meff=M\Phi_M(x)$ describes the decrease of $m^{(1)}$, and well beyond the freezing point the layer returns the training example closest to the query, $n_{\max}\to1$ and $m^{(1)}\to r$.

The generalization window \eqref{eq:window} is bounded by the crosstalk edge, where the noise coming from the other classes equals that coming from the finite number of examples, and by the condensation edge, beyond which the correct class is represented by a finite number of examples; the form \eqref{eq:window_emp} applies to any dataset once $\bar m$ and $\sigma_m$ are measured query by query. At the crosstalk edge the archetype is already retrieved ($m^{(1)}\simeq\erf(1/\sqrt{4\rho})$; at $\rho=0.2$ the full expression \eqref{eq:rhoeff} gives $\rho_{\rm eff}=0.398$, at $\rho=1$, where $x_{\rm lo}=1.4$, it gives $3.7$), well before the softmax has selected its class, because the weight lost to the $K-1$ wrong classes is spread over all of them (middle column of Fig.\ \ref{fig:lambda}: at $\beta_{\rm lo}$, $W=0.19$ and $m^{(1)}=0.89$ from \eqref{eq:main}, $0.86$ from the Monte Carlo). Above $\beta_{\rm lo}$ the crosstalk keeps decaying and $W$ saturates to one, while the pooling degrades as $x^2$; the magnetization is maximal where the two effects balance and then decreases through the window. The window opens as $\sqrt N$ (at fixed load $\ln K=\ln\alpha N$ subtracts a $\ln\ln N$), and the number of archetypes that can be learnt from $M$ examples of quality $r$, in the sense that the correct class is selected, grows as $\exp(c\sqrt N)$, with $c<c_{\rm crit}$ once the self-coupling is included (Remark \ref{rem:largeK}), faster than any power of $N$ but not exponentially in $N$; with or without the self-coupling, this condition is the exponential-model counterpart of the condition $M\gtrsim r^{-2}$ of \cite{Alemanno2023}.

\begin{figure}[t]
\centering
\includegraphics[width=\textwidth]{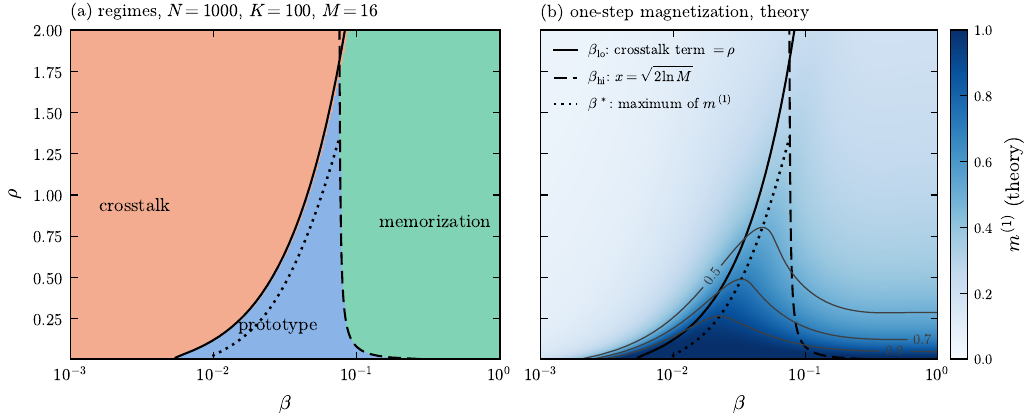}
\caption{The three regimes in the $(\beta,\rho)$ plane, from the closed-form theory alone ($N=1000$, $K=100$, $M=16$). (a) Crosstalk (second term of \eqref{eq:crosstalk} above $\rho$), prototype and memorization ($x>\sqrt{2\ln M}$) regions; solid and dashed lines are the edges \eqref{eq:window}; the dotted line is the position $\beta^*$ of the maximum of $m^{(1)}$, obtained by maximizing \eqref{eq:main} numerically and drawn where it lies below the condensation edge; above $\rho\simeq1.3$ the maximum crosses the condensation edge, where the Gaussian closure is no longer reliable. The crosstalk edge moves with $\rho$ as $1/r^2$, the condensation edge as $1/\sqrt{1-r^2}$, and the drawn edges close at $\rho\simeq1.8$, Eq.\ \eqref{eq:Kmax} (with the self-coupling the closure moves to $\rho\simeq1.4$, since $K>K_\times$ there); at large $\rho$ the pooling loss is already appreciable at $\beta_{\rm lo}$ ($x_{\rm lo}>1$), and the crosstalk edge marks the point where the crosstalk drops below $\rho$, not a fixed value of $m^{(1)}$. (b) One-step magnetization \eqref{eq:main}, with the contours $m^{(1)}=0.5$, $0.7$, $0.9$: the prototype regime is where the layer pools the correct class, not where it necessarily retrieves the archetype, since $\rho_{\rm eff}\to\rho$ there; in the memorization region \eqref{eq:main} overestimates $m^{(1)}$, whose actual value tends to $r$ (Remark \ref{rem:limits}).}
\label{fig:phase}
\end{figure}

\begin{remark}[range of validity]\label{rem:limits}
The Gaussian step of the derivation requires the field to be a sum of many comparable terms; it fails when the softmax has condensed on a few examples. The exact limit is then that of Remark \ref{rem:cond}, and Eq.\ \eqref{eq:main} overestimates $m^{(1)}$ by up to $0.13$ in that regime (Sec.\ \ref{sec:num}). The self-coupling $\tau\sigma^{(0)}_i$ is a binary term treated as Gaussian: where it dominates the field ($|h^{\setminus i}_i|\ll\tau$ with high probability, i.e.\ $rW\ll\tau$ and $\mathcal V-\tau^2(1-q^2)\ll\tau^2$) the update returns the query and $m^{(1)}=q$ exactly, while \eqref{eq:main} gives $\erf(q/\sqrt{2(1-q^2)})$ ($0.44$ at $q=0.5$); for $rW\simeq\tau$, as at the crosstalk edge for $K\gtrsim K_\times$, the field is a binary mixture, with $m^{(1)}\simeq(1+q)/2$ when the rest of the noise is small. Treating it exactly, i.e.\
\begin{equation*}
m^{(1)}=\sum_\pm\frac{1\pm q}{2}\erf\big[(rW\pm\tau)/\sqrt{2(\mathcal V-\tau^2(1-q^2))}\big],
\end{equation*}
changes $m^{(1)}$ by less than $0.01$ below $\beta_{\rm hi}$ at the parameters of Figs.\ \ref{fig:phase} and \ref{fig:lambda}, and leaves $c_{\rm crit}$ unchanged (Appendix D).
\end{remark}

\begin{remark}[optimal temperature]\label{rem:max}
By Proposition \ref{prop:opt} the optimum does not remove the crosstalk: at $\beta^*$ the crosstalk is a fraction $1/(N\beta^*r^2)\simeq1/\ln K$ of the pooling loss, Eq.\ \eqref{eq:balance}, because the former is exponentially cheap to reduce and the latter is not. The correct class then carries a weight $W^*<1$ ($0.75$, $0.50$ and $0.39$ for the three columns of Fig.\ \ref{fig:lambda}, against $0.74$, $0.49$, $0.39$ from the full expression \eqref{eq:rhoeff}), which decreases with the load: the layer generalizes best while it still shares its weight with the wrong classes. The maximum lies inside the window, at $\beta^*/\beta_{\rm lo}=1.3$--$1.8$ for the parameters explored here; Fig.\ \ref{fig:phase} draws it in the $(\beta,\rho)$ plane.
\end{remark}

\begin{remark}[the load disappears]
In \eqref{eq:rhoeff} the number of classes enters only through the crosstalk term, which for $q=r$ is $(1+\rho)(K-1)e^{-2N\beta r^2+N\beta^2r^2}$: the crosstalk of $\alpha N$ archetypes, which is extensive in the pairwise model and produces the spin-glass phase at $\alpha>\alpha_c(\rho)$, becomes exponentially small in $N\beta$ above $\beta_{\rm lo}$, of order $1/K$ at $2\beta_{\rm lo}$ and of order $e^{-c\sqrt N}$ near the condensation edge. This is the learning counterpart of the exponential storage capacity of \cite{Demircigil2017,Lucibello2024,Albanese2025exp}: the number of archetypes the exponential model can \emph{learn} from examples is limited not by the crosstalk but by the condensation edge (Remark \ref{rem:largeK}). In the dense $p$-body networks of \cite{Agliari2023sup,Agliari2023unsup} the crosstalk is still extensive in $\alpha_p=K/N^{p-1}$; here the polynomial load has become the logarithm in \eqref{eq:W}.
\end{remark}

\begin{remark}[overfitting as a REM transition]
In the pairwise model, overfitting is the spin-glass phase of a Hebbian spin glass with extensively many quenched patterns, a phase with full replica-symmetry breaking \cite{AGS1987,MPV1987,Albanese2022RSB}. In the exponential model, overfitting is the condensation of Proposition \ref{prop:rem}: the Gibbs measure of the attention layer freezes on one training example, as in the low-temperature phase of the REM, which is the $p\to\infty$ limit of the $p$-spin glass and has one step of replica-symmetry breaking \cite{Derrida1981,Gross1984}. We use the REM language for the softmax weights seen from a fixed query, which is where Lemma \ref{lem:logits} applies, and do not claim a thermodynamic phase transition of the network at finite $M$.
\end{remark}

\begin{remark}[class averages as keys]\label{rem:sup}
If the labels are disclosed and the class averages are used as keys, the within-class condensation cannot occur, since each class has a single key: by Proposition \ref{prop:sup} the prototype regime extends to $\beta\to\infty$, where the update is the majority rule of \cite{Alemanno2023} and the layer is a nearest-centroid classifier, i.e.\ it assigns the query to the class whose prototype $\bar\e^\mu$ has the largest overlap with it (for $K=1$, $\s^{(1)}=\sgn\bar\e^1$ for every $\beta$ and $q$). The interesting physics, and the transformer-relevant one, is in the keys $=$ examples.
\end{remark}

\section{Numerical results}\label{sec:num}
\begin{figure}[t]
\centering
\includegraphics[width=\textwidth]{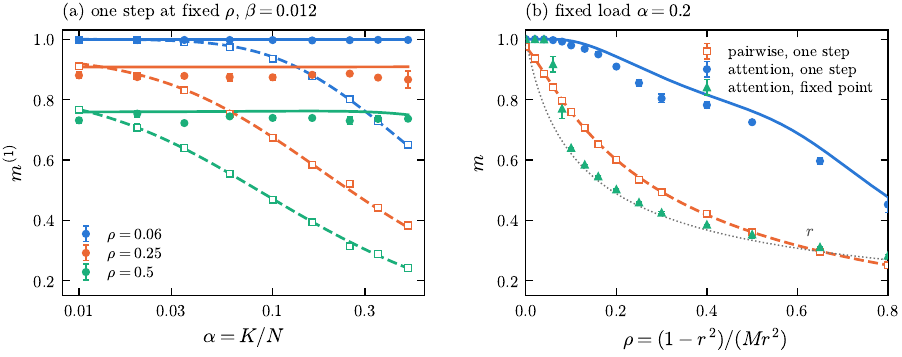}
\caption{The load disappears. Retrieval of an unseen archetype from a fresh test example ($q=r$), $N=4000$, $M=16$, $\beta=0.012$; lines are theory, symbols Monte Carlo (3 samples $\times$ 6 queries, standard errors). (a) One-step magnetization against the load $\alpha$ at three dataset entropies: the pairwise network, Eq.\ \eqref{eq:pair} (dashed lines, open squares), degrades as $\alpha$ grows; the attention layer, Eqs.\ \eqref{eq:main}--\eqref{eq:rhoeff} (solid lines, full circles), does not. (b) Magnetization against $\rho$ at $\alpha=0.2$: pairwise (open squares), attention after one step (full circles) and at the fixed point of the iterated map (triangles); the dotted line is the quality $r$ of a single example. The iterated map retrieves the archetype only for $\rho\lesssim0.06$ and otherwise converges to a training example.}
\label{fig:maps}
\end{figure}

\begin{figure}[t]
\centering
\includegraphics[width=\textwidth]{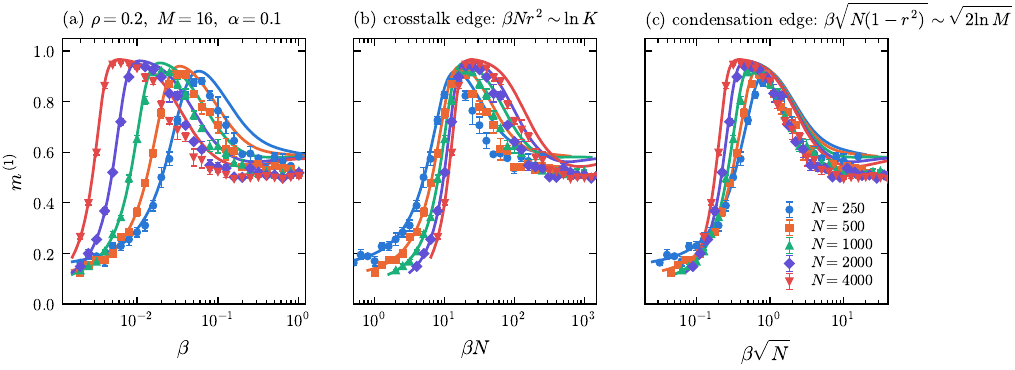}
\caption{Size dependence of the generalization window at $\rho=0.2$, $M=16$, $\alpha=0.1$ (lines: theory; symbols: Monte Carlo, 3 samples $\times$ 8 queries). (a) Against $\beta$ the window widens with $N$. (b) Against $\beta N$ the rising (crosstalk) edges collapse, up to the slow drift $\ln K=\ln(\alpha N)$ of Eq.\ \eqref{eq:window}. (c) Against $\beta\sqrt N$ the falling (condensation) edges collapse.}
\label{fig:N}
\end{figure}

\begin{figure}[t]
\centering
\includegraphics[width=0.62\textwidth]{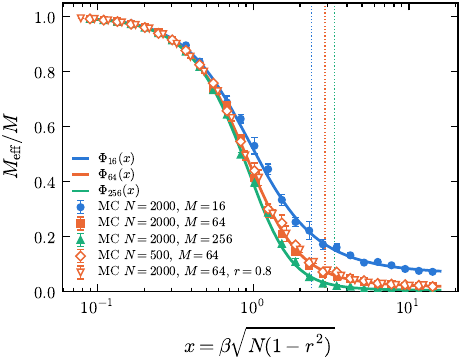}
\caption{Condensation of the within-class softmax. The pooled fraction $\Meff/M$ measured in simulations with different $N$, $M$ and $r$ (symbols; $r=0.5$ unless stated) collapses, as a function of the single variable $x=\beta\sqrt{N(1-r^2)}$, on the participation ratio $\Phi_M(x)$ of a random energy model with $M$ levels (lines, Proposition \ref{prop:rem}); dotted verticals mark the REM freezing point $x_c=\sqrt{2\ln M}$.}
\label{fig:rem}
\end{figure}

Simulations implement Definitions \ref{def:data}--\ref{def:dyn} literally: a dataset is drawn, the network is prepared in a fresh example of a randomly chosen archetype and updated in parallel at zero temperature; $m^{(1)}$, $n_{\max}$, $W$ and $\Meff$ are measured after the first update and $m^{(\infty)}$ after convergence (Appendix A). Each point is an average over independent datasets, each averaged over its queries; error bars are standard errors over the dataset averages, which account for the correlation of queries sharing a dataset. Theory curves contain no fitted parameter.

\paragraph{Independence from the load.} Figure \ref{fig:maps}(a) compares, at three dataset entropies, the one-step magnetization of the pairwise network built on class averages, Eq.\ \eqref{eq:pair}, with that of the attention layer at an inverse temperature inside the window \eqref{eq:window}, as the load is varied over almost two decades. The pairwise curves bend down with $\alpha$, as in \cite{Alemanno2023}: at $\rho=0.06$ the magnetization drops below $0.9$ at $\alpha\simeq0.14$ and reaches $0.65$ at $\alpha=0.5$; at $\rho=0.25$ it is below $0.9$ already at $\alpha\simeq0.015$ and reaches $0.38$; at $\rho=0.5$ it reaches $0.24$. The attention curves are flat over the whole range, at $1.0$, $0.90$ and $0.75$ respectively, which are their $\alpha\to0$ values, i.e.\ the majority-rule values $\erf(1/\sqrt{2\rho})$ lowered by the partial pooling $\Meff/M\simeq0.65$ at this $\beta$. The flatness does not require $W$ to saturate: at $\rho=0.5$ the correct class carries only $W\simeq0.3$ of the weight at $\alpha=0.1$ and $W\simeq0.1$ at $\alpha=0.5$, yet $m^{(1)}$ does not move, because the lost weight is spread over all the wrong classes: its contribution to the noise, the term $(1-W)^2/[(K-1)W^2]$ in \eqref{eq:rhoeff}, stays well below $\rho$ ($0.04$ at $W=0.1$, $K=2000$), whereas in the pairwise model the same crosstalk is of order $\alpha$. The dependence on the load returns only at the crosstalk edge of \eqref{eq:window}, where that term becomes comparable with $\rho$. Panel (b) shows the dependence on $\rho$ at the fixed load $\alpha=0.2$: the attention layer stays above the pairwise network at every $\rho$, and both follow their one-step theories. The triangles are the fixed point of the iterated attention map: for $\rho\lesssim0.06$ the archetype itself, which is not one of the keys, is an attractor of the softmax dynamics and is retrieved exactly; for larger $\rho$ the fixed point is a training example ($n_{\max}=1$, $m^{(\infty)}\simeq r$, the dotted line). Iterating a transformer layer memorizes unless the examples are good enough for the pooled majority vote to be self-consistent; a single application generalizes over a wider range. This is a departure from the attractor paradigm of associative memory: with examples rather than archetypes as keys, the archetype is reached by stopping after one step rather than by converging. It is also, we note, how attention is used in practice, where each layer is applied once in a feed-forward pass rather than iterated to a fixed point (iterated variants exist \cite{Hoover2024}); the present setting offers one reason why the single application need not be a loss.

\paragraph{Temperature scan.} Figure \ref{fig:lambda} follows the three regimes of Corollary \ref{cor:regimes}, anticipated on one query in Fig.\ \ref{fig:regimes}, as $\beta$ is varied. The magnetization is a non-monotonic function of $\beta$, with its maximum inside the window \eqref{eq:window}, at the $\beta^*$ of Proposition \ref{prop:opt} ($1.4\,\beta_{\rm lo}$ for the middle column), where the crosstalk has fallen well below $\rho$ while $\Meff$ is still close to $M$; the measured maxima, $1.00$, $0.92$ and $0.77$ for the three columns, approach the majority-rule values $\erf(1/\sqrt{2\rho})=0.998$, $0.975$ and $0.84$. Through the window the pooling degrades as $\Phi_M(x)$ and $m^{(1)}$ decreases; beyond the condensation edge $n_{\max}$ rises to one and $m^{(1)}$ settles at about $r$: the layer has become a lookup table of its training set. The closed-form theory reproduces $W$ and $\Meff$ quantitatively at all $\beta$ (within $0.04$), and $m^{(1)}$ while the pooling is substantial (within $0.04$ for $\Meff/M>0.9$, $0.06$ for $\Meff/M>0.5$); as the pooling degrades it overestimates $m^{(1)}$ increasingly, by up to $0.09$ for $\Meff/M>0.3$ and $0.13$ in the condensed regime, because the Gaussian approximation no longer holds when a few keys carry the weight (Remark \ref{rem:limits}). The measured plateaus, $0.67$, $0.53$ and $0.48$ for $r=0.62$, $0.49$, $0.45$ (left to right), lie slightly above $r$, the excess being the selection bias of the closest example.

\paragraph{Scaling of the two edges.} The two edges of the window have different scaling variables, and Fig.\ \ref{fig:N} isolates them: plotted against $\beta N$ the rising edges of the curves for $N=250$ to $4000$ collapse, plotted against $\beta\sqrt N$ the falling edges collapse. The window opens as $\sqrt N$, as predicted by \eqref{eq:window}; at $N=250$ it is a factor of four wide, at $N=4000$ a factor of ten.

\paragraph{The condensation is a REM transition.} Figure \ref{fig:rem} tests Proposition \ref{prop:rem} directly. Simulations at different $N$, $M$ and $r$, once plotted against $x=\beta\sqrt{N(1-r^2)}$, fall on the REM participation ratio $\Phi_M(x)$ with no adjustable parameter, including the $r=0.8$ series, whose variance $1-r^2$ differs from the others by a factor of two. The departure of $\Meff$ from $M$ sets in at $x\sim1$, as the small-$x$ expansion of Proposition \ref{prop:rem} says, and the condensation is essentially complete beyond $x_c=\sqrt{2\ln M}$, in the sense that the pooled fraction is small, $\Phi_M(x_c)=0.21$, $0.07$, $0.02$ for $M=16$, $64$, $256$, while $\Meff$ itself is still $3$--$6$ examples there and reaches $\simeq2$ at $2x_c$: the nearest key is returned only well beyond $x_c$ (in Fig.\ \ref{fig:lambda} $n_{\max}$ reaches one at $\beta\simeq2.3\,\beta_{\rm hi}$). The condensation edge of an attention layer is therefore not a property of the particular dataset but a universal feature of a softmax over $M$ exchangeable keys: $\beta_{\rm hi}$ depends on the data only through the conditional spread $\sqrt{(1-r^2)/N}$ of the within-class overlaps.

\begin{figure}[t]
\centering
\includegraphics[width=0.62\textwidth]{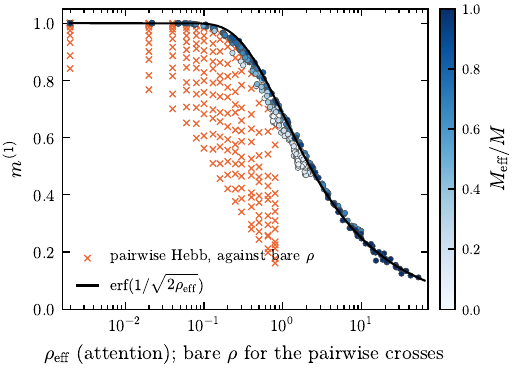}
\caption{All Monte Carlo points of Figs.\ \ref{fig:lambda} and \ref{fig:N} and of the $(\alpha,\rho)$ grid at $N=1000$ described in Appendix A, for the attention layer (every $\alpha$, $\rho$, $\beta$, $N$; colour: pooled fraction $\Meff/M$), plotted against the attention entropy $\rho_{\rm eff}$ evaluated from the measured weights, Eq.\ \eqref{eq:rhoeff} with the self-coupling to first order, $\tau\to\beta$; the line is $\erf(1/\sqrt{2\rho_{\rm eff}})$. Crosses: the pairwise network on the same grid, against the bare dataset entropy $\rho$, which does not account for the crosstalk.}
\label{fig:collapse}
\end{figure}

\paragraph{The attention entropy.} Figure \ref{fig:collapse} summarizes the content of Proposition \ref{prop:main}: when $\rho_{\rm eff}$ is evaluated from the softmax weights actually produced by each query, the one-step magnetization of every simulation, whatever $\alpha$, $\rho$, $\beta$ and $N$, lies on the single curve $\erf(1/\sqrt{2\rho_{\rm eff}})$. This is a test of the Gaussian closure given the attention statistics; the prediction of the statistics themselves by \eqref{eq:W} and \eqref{eq:Meff} is tested in Figs.\ \ref{fig:lambda} and \ref{fig:N}. Points in the condensed regime (light colours) lie below the curve, by up to $0.1$, where the Gaussian approximation fails (Remark \ref{rem:limits}). The pairwise network does not collapse against $\rho$, because the load is a second, independent source of noise there.

\section{Structured data}\label{sec:mnist}

\begin{figure}[t]
\centering
\includegraphics[width=\textwidth]{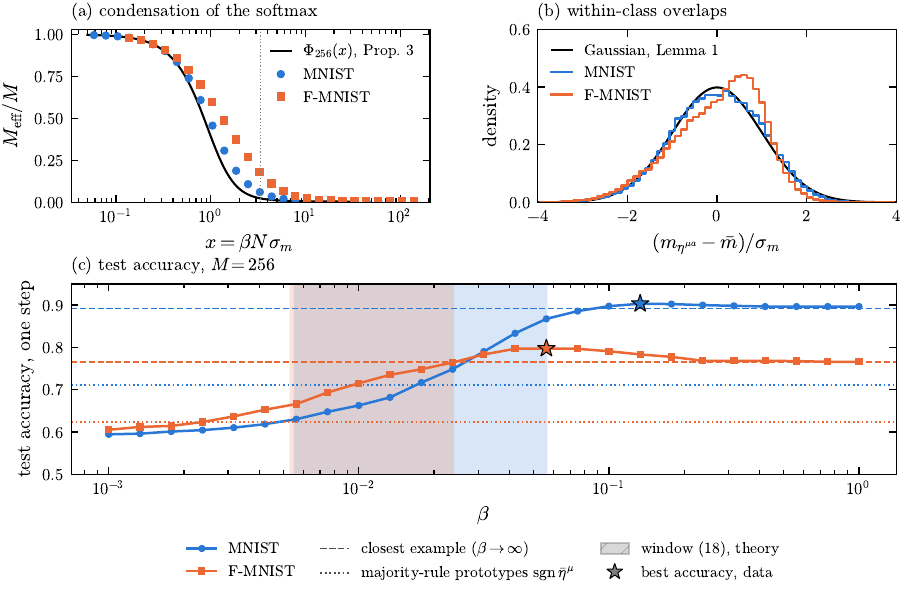}
\caption{One attention step with keys $=$ binarized MNIST and fashion-MNIST (F-MNIST) examples, $M=256$ per class, $K=10$, $N=784$, queries from the test sets (3 samples $\times$ 400 queries). (a) Pooled fraction $\Meff/M$ against $x=\beta N\sigma_m$, $\sigma_m$ being the measured standard deviation of the within-class overlaps; line: $\Phi_{256}(x)$ of Proposition \ref{prop:rem}; dotted vertical: $\sqrt{2\ln M}$. (b) Distribution of the standardized within-class overlaps against the Gaussian of Lemma \ref{lem:logits}. (c) Test accuracy of the one-step readout (class of largest total weight) against $\beta$; dashed: the $\beta\to\infty$ limit, which returns the closest training example; dotted: the nearest majority-rule prototype, $\arg\max_\mu\s\cdot\sgn\bar\e^\mu$; shaded: the window \eqref{eq:window_emp} with the measured $\bar m$ and $\sigma_m$ (and $\rho=N\sigma_m^2/(M\bar m)$); stars: the best accuracy over the $\beta$ grid, measured on the test queries.}
\label{fig:mnist}
\end{figure}

The theory of Sec.\ \ref{sec:snr} rests on Lemma \ref{lem:logits}: seen from the query, the within-class overlaps are independent Gaussian variables, so that the softmax is a REM. On real data the archetype is not available and the examples of a class are not independent blurred copies of one vector; in \cite{Alemanno2023} they were shown to be organized in a quasi-ultrametric way, with an overlap distribution reminiscent of replica-symmetry breaking, more markedly in fashion-MNIST than in MNIST. We therefore ask which parts of the picture survive on MNIST and fashion-MNIST (F-MNIST), binarized as in \cite{Alemanno2023} by thresholding each pixel at its mean over the training set, with the examples themselves as keys and the test images as queries; the labels enter only the readout (the class of largest total weight) and the accuracy, not the weights. The ingredients of the theory that can be measured on the data are the mean $\bar m$ and the standard deviation $\sigma_m$ of the overlaps between a query and the examples of its class. They enter the window in the form \eqref{eq:window_emp} and the condensation variable as $x=\beta N\sigma_m$; the parameter $\rho=N\sigma_m^2/(M\bar m)$ ($0.025$ on MNIST, $0.20$ on F-MNIST at $M=256$) is here an empirical quantity defined by the two measured moments, which coincides with the dataset entropy only for the data of Definition \ref{def:data}. Figure \ref{fig:mnist} collects the results for $M=256$ examples per class; $M=16$ and $64$ give the same picture with lower accuracies.

\paragraph{Condensation.} Panel (a) repeats the test of Fig.\ \ref{fig:rem} with $x=\beta N\sigma_m$. On MNIST the pooled fraction follows $\Phi_M(x)$ with no adjustable parameter, within $0.1$ around $x\simeq1$ and within $0.02$ beyond $x_c$, as on Rademacher data. On F-MNIST it condenses later and keeps a tail: $\Meff/M=0.11$ at $x=4.4$ and $0.023$ at $x=10$, where the REM gives $0.014$ and $0.006$. Panel (b) shows why: the standardized within-class overlaps of MNIST are Gaussian (skewness $-0.16$, excess kurtosis $0.0$), those of F-MNIST are not (skewness $-0.7$, excess kurtosis $1.2$), and in particular their upper tail is compressed, with $\mathcal P(z>2)=0.006$ against $0.023$ for a Gaussian and a gap between the two largest overlaps of $0.11\,\sigma_m$ against the $0.32\,\sigma_m$ expected for $256$ Gaussian levels. A query of F-MNIST has many nearly equivalent closest examples, and a softmax over nearly degenerate levels cannot freeze on one of them: this is what sub-clusters within the class would produce, consistent with the hierarchical organization found in \cite{Alemanno2023}, and it is where the REM should be replaced by its hierarchical generalization \cite{Derrida1985,DerridaGardner1986}. The diagnostic of Proposition \ref{prop:rem} thus not only measures memorization but also detects structure.

\paragraph{Accuracy.} Panel (c) shows the three regimes on the classification accuracy of the one-step readout. At small $\beta$ the readout tends to the nearest unnormalized class average, $\arg\max_\mu\s\cdot\bar\e^\mu$, which gives $0.6$ on both datasets, below the majority-rule prototypes. It then rises through the window \eqref{eq:window_emp} and reaches its maximum \emph{beyond} the condensation edge, at $\beta\simeq2.4\,\beta_{\rm hi}$ on MNIST ($0.90$, with $\Meff/M=0.01$) and $\beta\simeq2.3\,\beta_{\rm hi}$ on F-MNIST ($0.80$, with $\Meff/M=0.04$), far above the majority-rule prototype readout ($0.71$ and $0.62$). On MNIST the accuracy then stays at the value of the closest-example limit; on F-MNIST it is non-monotonic and its maximum exceeds that limit by $0.03$ (the best point of the test grid, within the fluctuations of three samples). The overlap of the output state with the class prototype behaves instead as on Rademacher data: it is maximal inside the window, at $\beta\simeq0.02$ on MNIST ($0.84$) and $\beta\simeq0.013$ on F-MNIST ($0.70$), and decreases on condensation. The two maxima are separated because on structured data the class prototype is a poor target: it is the replica-symmetric representation of the class, which in \cite{Alemanno2023} gave $75\%$ on MNIST against $95\%$ once a hidden layer of sub-prototypes was added. Here the same gain is obtained without a hidden layer, by the temperature: at $\beta$ a few times $\beta_{\rm hi}$ the softmax pools a few examples around the query, i.e.\ it reads out a local sub-prototype rather than the class average, and the memorization regime of the Rademacher theory becomes the useful operating point. In the language of Corollary \ref{cor:regimes}, the optimal $\Meff$ is not $M$ but the size of the sub-cluster to which the query belongs; on data with a single archetype per class that size is $M$, on hierarchical data it is of order one.

\section{Gradient descent without labels}\label{sec:gd}

\begin{figure}[t]
\centering
\includegraphics[width=\textwidth]{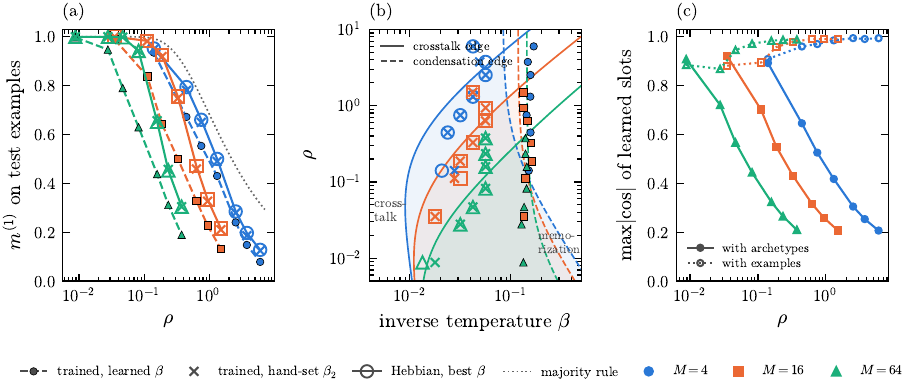}
\caption{Attention head with keys $=$ values and one slot per training example ($H=KM$), trained by gradient descent, without labels, on the denoising objective \eqref{eq:denoise} ($r'=0.5$), on the datasets of Definition \ref{def:data} ($N=400$, $K=20$). In (a) and (b) full symbols refer to the trained head at its learned inverse temperature, crosses to the trained head read out at a hand-set $\beta_2$, large open symbols to the Hebbian head with keys $=$ examples at its best $\beta$; colours and shapes distinguish $M$. (a) $m^{(1)}$ on fresh examples; dotted: majority-rule value $\erf(1/\sqrt{2\rho})$. (b) Inverse temperatures on the phase map of Eq.\ \eqref{eq:window} for each $M$ (solid: crosstalk edge; dashed: condensation edge; shaded: window); the learned one is in the units of the Hebbian head. (c) Largest cosine of the learned slots with the archetypes (full symbols) and with the training examples (open symbols), averaged over the slots with weights given by their mean attention on the test queries.}
\label{fig:gd}
\end{figure}

The Hebbian prescription sets the keys equal to the examples. We now ask what a head of the same form learns when its keys are trained by gradient descent, still without labels. The head has $H=KM$ learnable slots $\bm k_b$, one per training example, used both as keys and as values, $\bm h(\s)=\sum_{b=1}^Hw_b(\s)\,\bm k_b$ with the weights $w_b$ of Eq.\ \eqref{eq:attn} and a learnable inverse temperature $\beta$, and it is trained to reconstruct each training example from a corrupted copy of itself, by minimizing
\begin{equation}
\mathcal L=\frac{1}{KMN}\sum_{\mu=1}^K\sum_{a=1}^M\E_{\bm\chi'}\big\|\bm h(\e^{\mu a}\odot\bm\chi')-\e^{\mu a}\big\|^2,
\label{eq:denoise}
\end{equation}
where $\odot$ is the entrywise product and $\mathcal P(\chi'_i=\pm1)=(1\pm r')/2$, with $r'=0.5$; the average over $\bm\chi'$ is replaced by a fresh corruption at every step. Neither the labels nor the archetypes enter the objective. On fresh test examples we measure, as in Sec.\ \ref{sec:num}, the overlap $m^{(1)}$ of $\sgn\bm h$ with the unseen archetype, both at the learned inverse temperature and at a second one, $\beta_2$, set by hand after training. The latter is expressed in the units of Eq.\ \eqref{eq:attn}, i.e.\ with logits $\beta_2\sqrt N\,\bm k_b\cdot\s/\bar k$, $\bar k$ being the mean norm of the slots, and is chosen on the same grid and by the same criterion as the $\beta$ of the Hebbian head, namely the test overlap; both are thus oracle choices on the test examples, the same for the two heads, not a procedure for setting the temperature.

Figure \ref{fig:gd} collects the results. The learned slots stay on the training examples: their largest cosine with an example is $0.95$--$0.99$ for $r\le0.5$, while that with the archetypes is close to $r$, the overlap of a single example with its archetype (panel c). Gradient descent thus learns example-like keys and, functionally, rebuilds the Hebbian head (we have not checked that every example is covered by a slot). Panel (b) places the learned temperature, in the units of the Hebbian head ($\beta\bar k/\sqrt N$), on the phase map of Eq.\ \eqref{eq:window}: it lies near the condensation edge ($0.55$--$2.0$ times $\beta_{\rm hi}$, beyond it for $12$ of the $21$ datasets), $2.4$--$10$ times the best $\beta$ of the Hebbian head. The latter lies inside the window on $13$ of the $21$ datasets, at $1.0$--$2.1$ times $\beta_{\rm lo}$ as in Remark \ref{rem:max}, and just below the crosstalk edge on the noisiest ones, where the window is narrow or closed. The cost is that of the condensation side of the window, Corollary \ref{cor:regimes}(iii): at its learned temperature the trained head retrieves the archetype less well than the Hebbian head at its best $\beta$, by up to $0.31$ in $m^{(1)}$ (median $0.11$, panel a), and it never exceeds the majority-rule value. Read out at $\beta_2$, the same trained slots retrieve the archetype as well as the Hebbian head, within $0.013$ at every point, and the best $\beta_2$ coincides with the best Hebbian $\beta$ on $17$ of the $21$ datasets. On these data the denoising objective therefore learns the keys of the theory and misplaces only the temperature, in the direction that the phase diagram of Sec.\ \ref{sec:snr} identifies as memorization: an example is best reconstructed by attending to its own slot. These observations refer to this setup, one corruption level, full-batch Adam with a single learning-rate schedule and two datasets per point.

\section{Conclusions and outlook}

We have transferred Hebbian learning from examples \cite{Agliari2022concept,Alemanno2023} from the pairwise Hopfield network to the exponential one, whose zero-temperature update is a layer of softmax attention with the training examples as keys and values. The signal-to-noise analysis closes in this setting and yields a one-step magnetization $m^{(1)}=\erf(1/\sqrt{2\rho_{\rm eff}})$ that has the same majority-rule form as in the pairwise network but with the dataset entropy replaced by an attention entropy, Eq.\ \eqref{eq:rhoeff}, built from two analytically computable statistics of the softmax: the weight of the correct class and the number of examples it pools. The implications are of two kinds.

\paragraph{For Hopfield networks.} $(i)$ The load is no longer a control parameter of learning: within the window \eqref{eq:window} the crosstalk of the other archetypes decays exponentially in $N\beta$ instead of being of order $\alpha$, and the exponential network learns from examples a number of archetypes growing as $\exp(c\sqrt N)$, with
\begin{equation*}
c<c_{\rm crit}=r^2\sqrt{2\ln M/(1-r^2)}
\end{equation*}
(Remark \ref{rem:largeK}). The limit is set by the condensation edge rather than by the crosstalk, and at that growth rate the correct class is still selected but retrieved below its majority vote. The dense networks of \cite{Agliari2023sup,Agliari2023unsup}, with load $K/N^{p-1}$, are the intermediate steps of this family. $(ii)$ Overfitting changes nature. In the pairwise model it is a Hebbian spin glass with full replica-symmetry breaking; in the exponential model it is the condensation of the Gibbs measure over the keys onto one example, a random-energy transition with one step of replica-symmetry breaking at $\beta\sqrt{N(1-r^2)}=\sqrt{2\ln M}$, whose location depends on the data only through the spread of the within-class overlaps. The exponential model is, in this sense, the least glassy member of the Hebbian family, as the REM is of the $p$-spin family. $(iii)$ The dataset entropy generalizes to an entropy of the weighted vote: pairwise and exponential networks are the two ends of one formula, with $\Meff=M$ and extensive crosstalk on one side, $\Meff=M\Phi_M(x)$ and exponentially small crosstalk on the other. The labels are not needed: inside the window the examples of the correct class receive the weight by themselves ($W\to1$), and the attention layer, which never sees a label, performs the same majority vote as the pairwise network built on class averages at zero load, Eq.\ \eqref{eq:pair} with $\alpha=0$, up to the pooling loss (Fig.\ \ref{fig:maps}). $(iv)$ Iteration is not free. With keys equal to examples the fixed points of the dynamics are the examples, except at small $\rho$, where the archetype is an attractor; the retrieval of an unseen archetype is a one-step operation, not a convergence.

\paragraph{For attention layers.} $(i)$ An attention head over a set of noisy examples has a computable phase diagram in the parameters $(\rho,\beta,N,K,M)$, and the softmax temperature is the axis along which it moves from crosstalk to prototype to memorization; the two edges of the generalization window scale as $1/N$ and $1/\sqrt N$, so that a larger model generalizes from the same context over a wider range of temperatures. $(ii)$ Memorization is a measurable property of a head whose keys are training examples: the participation ratio of its weights drops to order one at the REM threshold, and its departure from the REM curve measures how far the data are from the single-archetype picture (Sec.\ \ref{sec:mnist}). $(iii)$ Gradient descent on a head of the same form, with one slot per example and trained without labels to denoise the examples, rebuilds the Hebbian head but sets the inverse temperature near the condensation edge, several times the optimum, and loses retrieval accordingly; read out at the Hebbian temperature it coincides with the Hebbian head (Sec.\ \ref{sec:gd}). The phase diagram thus locates a trained head and the price of its temperature. $(iv)$ The single application of a layer generalizes over a wider range than its fixed-point iteration (point $(iv)$ above); this is consistent with the feed-forward use of attention, although we do not claim it explains it. $(v)$ On structured data the memorization regime is not a failure but the operating point: the classes of MNIST and F-MNIST are made of sub-clusters, the class prototype is a poor target, and the softmax at a few times the condensation edge reads out the local sub-prototype, doing with the temperature what the hidden layer of \cite{Alemanno2023} did by construction.

These implications are structural rather than numerical: the data of the theory are unstructured, the layer is single, the dynamics is noiseless and the signal-to-noise theory is Gaussian, hence quantitative only while the softmax pools many examples (within $0.05$ with the measured weights, Fig.\ \ref{fig:collapse}) and off by up to $0.1$--$0.13$ once it has condensed. Several directions follow. The natural tool to close the gap in the condensed regime is the REM-like treatment of exponential models of \cite{Lucibello2024}, or the large-deviation evaluation of the noise of \cite{Agliari2026hetero}, with examples in place of archetypes; either would also give the selection bias of the closest example. The latter work, at $\beta\to\infty$, finds generalization to unseen cues at loads well below the memorization capacity: in the present picture that is the memorization regime, and the window of Corollary \ref{cor:regimes} is where the two capabilities meet. The iterated dynamics, in which the archetype emerges as a fixed point of the softmax map for small $\rho$ (Fig.\ \ref{fig:maps}b), deserves an analytic condition. The finite-temperature equilibrium statistics, which in the pairwise model gives the first-order retrieval transition of \cite{Alemanno2023}, is the obvious next step, as is the theory of structured datasets, where Sec.\ \ref{sec:mnist} indicates the route: a hierarchical random energy model \cite{Derrida1985,DerridaGardner1986} for the within-class overlaps, whose levels of condensation should correspond to the levels of the ultrametric organization uncovered in \cite{Alemanno2023}. Finally, the diagnostic of point $(ii)$ above and the denoising experiment of point $(iii)$ are the two statements that could be confronted with trained transformers, once the mapping from learned projections, where the logits are no longer the overlaps of Lemma \ref{lem:logits}, to the variable $x$ is understood.

\section*{Acknowledgments}
This work was supported by SolarSud T.R.E. di Miccoli Antonia. The research on this topic was accelerated by the use of Claude Fable 5.1 (Anthropic); the author takes full responsibility for the content. Code, data and the Monte Carlo results reproducing all figures are available at \url{https://github.com/francescoalemanno/hebb-attention}.

\appendix
\section{Details of the simulations}
Datasets are generated as in Definition \ref{def:data} with $r=1/\sqrt{1+\rho M}$. For the pairwise networks the field is $h_i=\sum_{j\ne i}J_{ij}\sigma_j$ with the couplings of Definition \ref{def:pair}; for the exponential network it is Eq.\ \eqref{eq:attn}, evaluated in log-space for numerical stability. A query is a fresh example of a randomly chosen archetype; the parallel update $\s\to\sgn(h)$ is applied once for $m^{(1)}$, $n_{\max}$, $W$, $\Meff$, and until a fixed point $\s^{(t+1)}=\s^{(t)}$ is reached, at most ten parallel updates, for $m^{(\infty)}$ (in a check with $60$ updates at the parameters of Fig.\ \ref{fig:maps}b no cycle occurred and $98\%$ of the trajectories reached their fixed point within ten updates, the rest within $13$, all at $\rho\simeq0.06$--$0.08$). Fig.\ \ref{fig:maps}: $N=4000$, $M=16$, $\beta=0.012$; (a) $9$ loads $\alpha\in[0.01,0.5]$ at $\rho=0.06,0.25,0.5$, (b) $15$ entropies $\rho\in[0,0.8]$ at $\alpha=0.2$; $3$ datasets $\times$ $6$ queries per point. Fig.\ \ref{fig:collapse} also uses a grid at $N=1000$, $\beta=0.025$: $13$ loads $\alpha\in[0.01,0.5]$ $\times$ $15$ entropies $\rho\in[0,0.8]$, $4$ datasets $\times$ $8$ queries per point. Fig.\ \ref{fig:regimes}: one dataset and one query, $N=1000$, $K=20$, $M=16$, $\rho=0.2$; the three values of $\beta$ are fixed by the edges \eqref{eq:window}. Fig.\ \ref{fig:phase}: theory only, $N=1000$, $K=100$, $M=16$, $240\times220$ grid in $(\beta,\rho)$. Fig.\ \ref{fig:lambda}:
\begin{gather*}
N=1000,\qquad K=100,\qquad (\rho,M)=(0.05,32),\,(0.2,16),\,(0.5,8),\\
28\text{ values of }\beta\in[2\cdot10^{-3},1],
\end{gather*}
$4\times8$ samples per point. Fig.\ \ref{fig:N}: $K=N/10$, $M=16$, $3\times8$ samples. Fig.\ \ref{fig:rem}: $K=2$, $N\in\{500,2000\}$, $M\in\{16,64,256\}$, $r\in\{0.5,0.8\}$, $4\times8$ samples; the abscissa is computed from $\beta$ as $x=\beta\sqrt{N(1-r^2)}$. The REM function $\Phi_M(x)$ is evaluated by averaging over $4000$ realizations of $M$ Gaussian levels. Fig.\ \ref{fig:mnist}: MNIST and F-MNIST binarized by thresholding each pixel at its mean over the $60{,}000$ training images; for each of $3$ samples, $M$ examples per class are drawn from the training set and $400$ queries from the test set; $25$ values of $\beta\in[10^{-3},1]$; $\bar m$, $\sigma_m$, skewness and kurtosis are computed query by query over the examples of the query's class, then averaged. Fig.\ \ref{fig:gd}: $N=400$, $K=20$, $H=KM$ slots initialized with i.i.d.\ Gaussian entries of variance $1/N$, $\beta$ initialized at $1/\sqrt N$, objective \eqref{eq:denoise} with $r'=0.5$ and a fresh corruption at every step, $1000$ full-batch Adam steps with learning rate $0.015$ for the first $500$ steps and $0.005$ after, $400$ fresh test examples, $2$ datasets per point; the Hebbian head is evaluated on the same data by one step of \eqref{eq:attn} with keys $=$ training examples; its $\beta$ and the readout temperature $\beta_2$ of the trained head are both chosen among $25$ values in $[10^{-3},1]$ on the test overlap, an oracle choice, as are the starred best accuracies of Fig.\ \ref{fig:mnist}. Monte Carlo check quoted in the proof of Proposition \ref{prop:main}: the annealed factor $e^{\beta^2Nr^2}[1+(e^{x^2}-1)/M]$ is $1.053$ and $1.124$ at $\beta=0.0134$ and $0.02$ for $N=1000$, $K=100$, $\rho=0.2$, $M=16$, and the measured ratio $(K-1)\sum_{\mu>1}V_\mu^2/(1-W)^2$ agrees with it.

\section{Small-$x$ expansion of $\Phi_M$}
Let $S_1=\sum_az_a$ and $S_2=\sum_az_a^2$. Expanding $e^{xz}=1+xz+\frac12x^2z^2+O(x^3)$ in the numerator and in the denominator of the participation ratio gives, realization by realization,
\begin{equation*}
\frac{(\sum_ae^{xz_a})^2}{\sum_ae^{2xz_a}}=M+x^2\Big(\frac{S_1^2}{M}-S_2\Big)+O(x^3);
\end{equation*}
the $O(x^3)$ term is a cubic polynomial in the $z$'s and averages to zero. Taking the average with $\E S_1^2=\E S_2=M$ gives
\begin{equation*}
M\Phi_M(x)=M-(M-1)x^2+O(x^4),
\end{equation*}
i.e.\
\begin{equation*}
\Phi_M(x)=1-(1-M^{-1})x^2+O(x^4);
\end{equation*}
numerically $\Phi_{16}(0.3)=0.921$, against $0.916$ from the expansion. For $x>x_c$ and $M\to\infty$, $\E\sum_aw_a^2=1-x_c/x$ in the frozen phase \cite{Derrida1985b,MPV1985}, and Jensen's inequality gives
\begin{equation*}
M\Phi_M(x)=\E[1/\sum_aw_a^2]\ge(1-x_c/x)^{-1};
\end{equation*}
the inequality is strict because $\sum_aw_a^2$ keeps a non-degenerate distribution. At $x=10$ the bound is $1.31$, $1.40$, $1.50$ for $M=16$, $64$, $256$, while $2\cdot10^4$ samples of $M$ Gaussian levels give $\E\sum_aw_a^2=0.82$, $0.77$, $0.72$ (above $1-x_c/x=0.76$, $0.71$, $0.67$, a finite-$M$ effect), hence $1/\E\sum_aw_a^2=1.21$, $1.30$, $1.39$, and $M\Phi_M=1.34$, $1.50$, $1.65$, i.e.\ $2$--$10\%$ above the asymptotic bound, and $11$--$19\%$ above $1/\E\sum_aw_a^2$ at the same $M$, which is the Jensen gap proper. For $M=16$, $\E[1/\Meff]$ exceeds $1/\E\,\Meff$ by $0.1\%$, $9\%$ and $27\%$ at $x=0.3$, $1$ and $2$; using it in \eqref{eq:rhoeff} lowers $m^{(1)}$ by less than $0.002$ at $\beta_{\rm lo}$, by $0.01$--$0.015$ at $\sqrt{\beta_{\rm lo}\beta_{\rm hi}}$ and by $0.04$--$0.06$ at $\beta_{\rm hi}$ for the parameters of Fig.\ \ref{fig:lambda}. In the text $\Phi_M$ is always evaluated numerically.

\section{Position of the maximum}
Write $t=(1-W)/W$, $u=(1-r^2)/\Meff$, $b=(r^2+u)/(K-1)$, $s=q\tau$ and $v=(1-q^2)\tau^2$; then \eqref{eq:rhoeff} reads
\begin{equation*}
\rho_{\rm eff}=\frac{u+bt^2+v(1+t)^2}{[r+s(1+t)]^2},
\end{equation*}
with $t=(K-1)e^{-N\beta qr+N\beta^2r^2/2}$ from \eqref{eq:W}, so that $t'=-Lt$ with $L=N(qr-\beta r^2)$, and $u=(1-r^2)/(M\Phi_M(x))$ from \eqref{eq:Meff}. The condition $\partial_\beta\rho_{\rm eff}=0$ is
\begin{align}
&[r+s(1+t)]\,[u'+b't^2+v'(1+t)^2]-2s'(1+t)\,[u+bt^2\nonumber\\
&\quad+v(1+t)^2]-2Lt\,\{[b(r+s)+vr]\,t+vr-us\}=0,
\label{eq:stat}
\end{align}
a scalar equation in $\beta$, exact within the signal-to-noise closure, that can be solved once $\Phi_M$ is known. For $x\ll1$,
\begin{equation*}
u\simeq u_0[1+(1-M^{-1})N(1-r^2)\beta^2]
\end{equation*}
with $u_0=(1-r^2)/M$; neglecting the self-coupling ($s=v=0$), the quadratic term in the exponent of $t$, and $\beta r$ with respect to $q$ in $L$, \eqref{eq:stat} reduces to $u'=2Lbt^2$, i.e.
\begin{equation}
(K-1)e^{-2Nqr\beta}=\kappa\beta,\qquad \kappa=\frac{(M-1)(1-r^2)^2}{M^2qr\,[r^2+(1-r^2)/M]},
\end{equation}
which for $q=r$ is $\kappa=(M-1)\rho^2/(1+\rho)$ and whose solution is
\begin{equation}
\beta^*=\frac{\mathcal W\big(2Nqr(K-1)/\kappa\big)}{2Nqr},\quad \frac{1-W^*}{W^*}=\sqrt{\kappa(K-1)\beta^*},
\label{eq:betastar}
\end{equation}
with $\mathcal W$ the Lambert function, $\mathcal W(y)e^{\mathcal W(y)}=y$. Equation \eqref{eq:balance} is the same condition written for the two terms of $\rho_{\rm eff}$, since $2Nqr\beta^*=2\ln[(K-1)/t^*]$; with $\beta^*\simeq\ln K/(Nr^2)$ inside the square root,
\begin{equation*}
\frac{1-W^*}{W^*}\simeq\sqrt{\alpha\ln K\;\frac{(M-1)\rho^2}{(1+\rho)\,r^2}}\,,
\end{equation*}
the load dependence quoted in Proposition \ref{prop:opt}. The expansion assumes $x^*\ll1$, i.e.\ a maximum well below the condensation edge, and is not uniform for $\rho\to0$, where $\kappa\to0$: at $N=1000$, $K=100$, $M=16$ the closed form is within $2\%$ of the numerical maximum of \eqref{eq:main} for $0.02\le\rho\le0.4$ ($x^*\le0.9$), $4\%$ off at $\rho=0.5$, $12\%$ at $\rho=0.8$ and $19\%$ at $\rho=1$, where $x^*>1$ (Fig.\ \ref{fig:phase}). The self-coupling shifts the maximum only slightly: minimizing the full expression \eqref{eq:rhoeff} at fixed load $\alpha=0.1$, $\rho=0.2$, $M=16$ gives $W^*=0.49$, $0.43$, $0.35$, $0.32$ for $N=10^3$, $10^4$, $10^6$, $10^7$, against $0.50$, $0.46$, $0.40$, $0.37$ from \eqref{eq:betastar}, with $\beta^*$ within $2\%$. Asymptotically the term in $s\,t$ of \eqref{eq:stat} takes over and $W^*$ decreases as $1/\ln N$ rather than $1/\sqrt{\ln N}$, but at any practical size \eqref{eq:betastar} is accurate.

\section{Crosstalk edge with the self-coupling}
With $\Meff=\Meff'=M$, $q=r$, $\epsilon=(1+\rho)/(K-1)$ and $\tau=\tanh\beta$, Eq.\ \eqref{eq:rhoeff} reads
\begin{equation*}
\rho_{\rm eff}=[\rho W^2+\epsilon(1-W)^2+M\rho\tau^2]/(W+\tau)^2,
\end{equation*}
and $\rho_{\rm eff}=2\rho$, the condition defining the crosstalk edge, is a quadratic equation in $W$ whose positive root is
\begin{equation*}
W_{\rm lo}=\frac{\epsilon+(M-2)\rho\tau^2}{\epsilon+2\rho\tau+\sqrt{(\epsilon+2\rho\tau)^2+(\rho-\epsilon)[\epsilon+(M-2)\rho\tau^2]}};
\end{equation*}
the edge is where $W(\beta)$ of \eqref{eq:W} reaches $W_{\rm lo}(\tanh\beta)$. At $\tau=0$, $W_{\rm lo}=\sqrt\epsilon/(\sqrt\epsilon+\sqrt\rho)$, which gives \eqref{eq:window}. For $\epsilon\ll\rho$ the discriminant is $\simeq\rho\epsilon+(M+2)\rho^2\tau^2$, so that the self-coupling is negligible as long as $(M+2)\rho\tau^2\ll\epsilon$, i.e.\ $K\ll K_\times$ with $\tau\simeq\beta_{\rm lo}$ (Remark \ref{rem:largeK}). In the opposite limit $\epsilon\to0$, $W_{\rm lo}\to(\sqrt{M+2}-2)\,\tau$, only polynomially small in $N$, and the edge becomes $N\beta_{\rm lo}r^2=\ln K-O(\ln N)$, twice \eqref{eq:window}. The constant $\sqrt{M+2}-2$ is that of the Gaussian closure; treating the binary self-coupling exactly (Remark \ref{rem:limits}), with the same condition $m^{(1)}=\erf(1/\sqrt{4\rho})$, changes it ($W_{\rm lo}\to3.1\,\tau$ instead of $2.2\,\tau$ at $M=16$, $\rho=0.2$) but not the proportionality to $\tau$, hence neither the $O(\ln N)$ term nor $c_{\rm crit}$. For $K=\exp(c\sqrt N)$, comparing with $\beta_{\rm hi}$ gives
\begin{equation*}
c_{\rm crit}=r^2\sqrt{2\ln M/(1-r^2)}=\sqrt{2\ln M/[\rho M(1+\rho M)]};
\end{equation*}
solving $W(\beta)=W_{\rm lo}(\tanh\beta)$ numerically with \eqref{eq:W}, the window closes at $c/c_{\rm crit}=1.027$, $1.006$, $1.001$, $1.000$ for $N=10^4$, $10^6$, $10^8$, $10^{10}$ ($M=16$, $r=0.5$). At the sizes of the figures the self-coupling moves $\beta_{\rm lo}$ by a few percent (by $-2.4\%$ at $N=1000$, $K=100$, $M=16$, $\rho=0.2$), except near the closure of the window in Fig.\ \ref{fig:phase}.

\end{document}